\documentstyle[11pt,epsf]{article}
\newcommand{\be}{\begin{equation}}
\newcommand{\ee}{\end{equation}}
\newcommand{\bee}{\begin{eqnarray}}
\newcommand{\eee}{\end{eqnarray}}

\newcommand{\sect}[1]{\section{#1}\setcounter{equation}{0}}
\newcommand{\slip}{\mathop{\sum\nolimits'}\limits}
\newcommand{\plip}{\mathop{\prod\nolimits'}\limits}
\begin{document}


\begin{center}
{\bf CALCULATION METHOD FOR THE THREE-DIMENSIONAL ISING 
FERROMAGNET THERMODYNAMICS WITHIN THE FRAMES OF $\rho^6$ MODEL}
\end{center}

\begin{center}
{\sc M.P.Kozlovskii, I.V.Pylyuk, V.V.Dukhovii}
\end{center}

\begin{center}
{\it Institute for Condensed Matter Physics \\
of the Ukrainian National Academy of Sciences, \\
1~Svientsitskii St., UA-290011 Lviv, Ukraine} \\
E-mail: piv@icmp.lviv.ua
\end{center}

\vspace{0.5cm}

{\small
Calculation of thermodynamic functions of the three-dimensional 
Ising ferromagnet above and below critical temperature is performed in the 
approximation of sixfold basis distribution ($\rho^6$ model). 
Comparison with the results for the $\rho^4$ model indicates that 
dependence of the thermodynamic functions on the renormalization group 
parameter $s$ becomes weaker. The optimal interval of the 
renormalization group parameter values is determined.
}

\vspace{0.5cm}

\section*{Introduction}

\noindent
Significant results in the description of the system thermodynamic 
properties in the vicinity of the transition point have been obtained by 
means of the collective variables (CV) approach. The method to deriving 
explicit expressions for the thermodynamic and correlation functions of the 
three-dimensional Ising model at temperatures both above and below critical 
temperature $T_c$ has been suggested within this approach. The calculations 
are performed with a non-Gaussian measure density. The measure is 
represented as an exponential function of the CV, the argument of which 
contains, along with the quadratic term, higher powers of the 
variable with the corresponding interaction constants. The simplest 
non-Gaussian measure density is the quartic one ($\rho^4$ model) with the 
second and the fourth powers of the variable in the exponent. Then the 
sixfold measure goes containing the sixth power of the variable ($\rho^6$ 
model), and so forth.

The results of the theory depend on the renormalization group (RG) parameter 
$s$ due to an approximation of the Ising model partition function calculation
using the non-Gaussian measure densities. This dependence decreases 
essentially if the non-Gaussian measure density becomes more complicated.
Calculations of the correlation length critical exponent $\nu$ within the 
$\rho^{2m}$ models with $m=2,3,4,5$ confirm this statement [1-3]. It has 
been established that the $\rho^6$ model provides an adequate description of 
the Ising model critical behaviour, in particular, the critical exponents, 
at the RG parameter values in the interval $2\leq s\leq 4$.
	
Investigation of the $\rho^6$ model within the numerical realization of 
the CV method has been performed in [4]. Analytical derivation of the 
explicit expressions for the $\rho^6$ model thermodynamic functions is the 
subject of the present paper. The foundations for such kind of 
investigations have been developed in [5-9], where the quartic 
distribution was used as a basis measure.

\sect{General relations}

The partition function of the three-dimensional Ising model within the 
sixfold measure density is given by
\bee
&& Z=2^N\int\exp [\frac{1}{2}\sum_{k\leq B}\beta\tilde{\Phi}(k)\rho_{\vec k}
	\rho_{-\vec k}+2\pi i \sum_{k\leq B}\omega_{\vec k}\rho_{\vec k}+ 
	{\sum_{n=1}}^3{(2\pi i)}^{2n} \times \nonumber \\
&&\times N^{1-n}\frac{1}{(2n)!}\sum_{k_1,
\ldots,k_{2n}\leq B} {\cal M}_{2n} \omega_{\vec k_1}\cdots \omega_{\vec 
k_{2n}}\delta_{\vec k_1+\cdots+\vec k_{2n}}](d\omega)^N(d\rho)^N,  
\eee
where ${\cal M}_2$=1, ${\cal M}_4$=-2, ${\cal M}_6$=16, $\tilde{\Phi}(k)=
\tilde{\Phi}(0)(1-2b^2k^2)$, $\beta = (kT)^{-1}$ is the inverse temperature, 
$b$ is the effective interaction radius of the potential $\Phi (r)=A\exp 
(-r/b)$, $\tilde{\Phi}(0)=8\pi A{(b/c)}^{3}$. Integrating (1.1) over 
$\rho_{\vec k}$ and $\omega_{\vec k}$ with the indices $B'<\mid\vec 
k\mid\leq B$ $(B=\pi/c, c$ is the simple cubic lattice constant), we get an 
expression for the partition function of the $\rho^6$ model:
\bee
Z&=&2^N 2^{\frac{N'-1}{2}} e^{{a'}_0 N'} \int\exp [-\frac{1}{2}
	\sum_{k\leq B'}d'(k)\rho_{\vec k}\rho_{-\vec k} - \\  
&& -\sum_{l=2}^3\frac{1}{(2l)!}(N')^{1-l}
	\sum_{k_1,\ldots ,k_{2l}\leq B'} a_{2l}^{'} \rho_{\vec k_1} \cdots
\rho_{\vec k_{2l}}\delta_{\vec k_1+\cdots+\vec k_{2l}}](d\rho)^{N'}. \nonumber 
\eee
Here $N'=N{s_0}^{-3}$, $s_0=B/B'=\pi\sqrt{2}b/c$,
\be
d'(k)=a_2^{'}-\beta\tilde{\Phi}(k).
\ee
Coefficients $a_{2l}^{'}$ depend on the ratio $b/c$ and are given 
by the relations
\bee
a_0^{'} &=& \ln Q({\cal M}),~~~~~Q({\cal M})=(12 s_0^3)^{1/4}\pi^{-1}
I_0(\eta',\xi'),\nonumber \\
a_2^{'} &=& (12s_0^3)^{1/2}{\cal F}_2(\eta',\xi'), \\
a_4^{'} &=& 12s_0^3 C(\eta',\xi'), \nonumber \\
a_6^{'} &=& (12s_0^3)^{3/2} N(\eta',\xi'), \nonumber
\eee
where the quantities $\eta'=\sqrt{3}s_0^{3/2}$, 
$\xi'=\frac{8\sqrt{3}}{15s_0^{3/2}}$ are the arguments, and special 
functions $C(\eta',\xi')$  and $N(\eta',\xi')$ read
\bee
C(\eta',\xi')  &=& -{\cal F}_4(\eta',\xi')+
3{\cal F}_2^2(\eta',\xi'),\\
N(\eta',\xi')  &=& {\cal F}_6(\eta',\xi')-15{\cal F}_4(\eta',\xi')
{\cal F}_2(\eta',\xi')+30{\cal F}_2^3(\eta',\xi'). \nonumber 
\eee  
Here ${\cal F}_{2l}(\eta',\xi')=\frac{I_{2l}(\eta',\xi')}{I_0(\eta',\xi')}$,
$I_{2l}(\eta',\xi')=\int_{0}^{\infty}t^{2l}e^{-\eta' t^2-t^4-\xi' t^6}dt$.

Using the method of layer-by-layer integration of the partition function in 
the phase space of CV, developed in [10], one can reduce (1.2) to the 
form:
\be
Z=2^N 2^{\frac{N_{n+1}-1}{2}} Z_0Z_1\ldots Z_n(Q(P_n))^{N_{n+1}}
\int w_6^{(n+1)}(\rho)(d\rho)^{N_{n+1}},
\ee
where $N_n=N's^{-3n}$,
\bee
&&Z_0=[Q({\cal M})Q(d)]^{N'},~~Z_1=[Q(P)Q(d_1)]^{N_1},~~\ldots~~,
	\nonumber \\ 
&&Z_n=[Q(P_{n-1})Q(d_n)]^{N_n}, \\
&&Q(P_n)=\frac{1}{\pi}(s^3\frac{a_4^{(n)}}{C(h_n,\alpha_n)})^{1/4}
	I_0(\eta_n,\xi_n), \nonumber \\
&&Q(d_n)=2(24/a_4^{(n)})^{1/4}I_0(h_n,\alpha_n). \nonumber
\eee
Hereafter, the arguments $h_n$, $\alpha_n$ are called basic:
\be
h_n=d_n(B_{n+1},B_n)(6/a_4^{n})^{1/2},
~~~\alpha_n=\frac{\sqrt{6}}{15}a_6^{(n)}/(a_4^{(n)})^{3/2}.
\ee
The effective measure density of the n-th phase layer $w_6^{(n)}(\rho)$
has the form:
\bee
w_6^{(n)}(\rho) &=& \exp[-\frac{1}{2}\sum_{k\leq B_n}d_n(k)\rho_{\vec k}
\rho_{-\vec k} - \\  
&-&\sum_{l=2}^3\frac{1}{(2l)!}N_n^{1-l}
\sum_{k_1,\ldots,k_{2l}\leq B_n} a_{2l}^{(n)}\rho_{\vec k_1} \cdots
\rho_{\vec k_{2l}}\delta_{\vec k_1+\cdots+\vec k_{2l}}]. \nonumber 
\eee
Here $B_n=B's^{-n}$. The intermediate variables $\eta_n,\xi_n$
are the functions of $h_n$ and $\alpha_n$: 
\bee
\eta_n &=& \sqrt{6}s^{3/2}{\cal F}_2(h_n,\alpha_n)[C(h_n,
	\alpha_n)]^{-1/2}, \\
\xi_n &=& \frac{\sqrt{6}}{15}s^{-3/2}N(h_n,\alpha_n)[C(h_n,
	\alpha_n)]^{-3/2}. \nonumber
\eee
The form of the special functions $C(h_n,\alpha_n),N(h_n,\alpha_n)$ is 
being given by (1.5).

Coefficients $d_n(B_{n+1},B_n), a_4^{(n)}, a_6^{(n)}$ are related 
to the coefficients of the $n+1$-th layer by the recurrent relations (RR) 
[11-13]. The solutions of these relations [13] are used in the 
calculation of the system thermodynamic characteristics.

\sect{Thermodynamic functions of the $\rho^6$ model in the regions of 
critical and limit Gaussian re\-gi\-mes (CR and LGR) above $T_c$}

It is convenient to rewrite the model partition function as [14]
\be
Z=2^NZ_{CR}Z_{LGR}.
\ee
Let us consider $Z_{CR}$ given by
\bee
Z_{CR}=\prod_{n=0}^{m_{\tau}}[\frac{2}{\pi}(\frac{24}{
C(\eta_{n-1},\xi_{n-1})})^{1/4}I_0(h_n,\alpha_n)I_0(\eta_{n-1},
\xi_{n-1})]^{N_n}.
\eee
It should be mentioned that in (2.2) $\eta_{-1}\equiv\eta',~~\xi_{-1}
\equiv\xi'$ at $n=0$. We represent the right-hand side (RHS) of (2.2) in the 
form of an explicit dependence on the phase layer number $n$ in order 
to calculate $Z_{CR}$.

In the CR region, the basic $h_n,\alpha_n$ and intermediate 
$\eta_n,\xi_n$ arguments are close to their values at the fixed point. 
Therefore, functions of these arguments can be written as power series of 
deviations of basic arguments from their values at the fixed point (see 
[15,16]). Using the obtained representations for $I_0(h_n,\alpha_n)$,
$I_0(\eta_{n-1},\xi_{n-1})$, $C(\eta_{n-1},\xi_{n-1})$, we 
determine from (2.2) the partial free energy corresponding to the $n$-th 
phase layer:
\bee
F_n &=& -kTN_n\{ 
	f_{CR}^{(0)}+\varphi_1(h_{n-1}-h^{(0)})+\varphi_2(\alpha_{n-1}-
	\alpha^{(0)})+ \nonumber \\
&&+\varphi_3(h_n-h^{(0)}) 
	+\varphi_4(\alpha_n-\alpha^{(0)})+\varphi_1^{'}(h_{n-1}-h^{(0)})^2+
	\nonumber\\
&&+\varphi_2^{'}(\alpha_{n-1}-\alpha^{(0)})^2+\varphi_3^{'}(h_n-h^{(0)})^2+ 
	\varphi_4^{'}(\alpha_n-\alpha^{(0)})^2+ \nonumber \\
&&+\varphi_5^{'}(h_{n-1}-h^{(0)})
	(\alpha_{n-1}-\alpha^{(0)})+\varphi_6^{'}(h_n-h^{(0)})(\alpha_n-
	\alpha^{(0)})\},
\nonumber \\
f_{CR}^{(0)}&=&\ln(\frac{2(24)^{1/4}}{\pi})-\frac{1}{4}\ln{\cal P}_{40}+
	\ln I_0^* +\ln I_0^{**}, \\
\varphi_m&=&b_m+{\cal P}_{4m}/4,~~ m=1,2, \nonumber \\
\varphi_3&=&-{\cal F}_2^*,~~ \varphi_4=-{\cal F}_6^*, \nonumber \\
\varphi_m^{'}&=&b_m^{'}-\frac{1}{2}b_m^2-{\cal P}_{4m}^{'}/4+{\cal P}_{4m}^2/8,~~
\varphi_3^{'}={\cal F}_{4}^*/2-{\cal F}_2^{*2}/2,   \nonumber \\
\varphi_4^{'}&=&{\cal F}_{12}^{*}/2-{\cal F}_6^{*2}/2,~~
\varphi_5^{'}=b_3^{'}-b_1 b_2-{\cal P}_{43}^{'}/4+{\cal P}_{41}{\cal P}_{42}/4,
	\nonumber \\
\varphi_6^{'}&=&{\cal F}_{8}^{*}-{\cal F}_2^*{\cal F}_6^*.   \nonumber
\eee
Expressions for the quantities occurring in $f_{CR}^{(0)}$, $\varphi_i$, 
$\varphi_j^{'}$ are given in [15,16].

Hence, the partial free energy of the $n$-th phase layer $F_n$ is 
written as a power series of deviations of basic arguments from their 
fixed point values. The linear approximation for $F_n$ was used in  [14]. 
In the present paper, as well as in the calculations within the $\rho^4$  
model [5,6], the quadratics of the deviations are also taken into 
account. It allows one to compare the results of the calculations for  
$\rho^4$ and $\rho^6$ models. Let us note that quadratic terms of the RR 
do not contribute to the elements of the matrix of the RR 
linearization in the vicinity of the fixed point, and the eigenvalues 
$E_l$ of this matrix and the critical exponent of the correlation length 
are the same as within the linear approximation for the RR.

Let us find an explicit dependence of $F_n$ on the layer number $n$. 
Using the solutions of the RR, we get for $h_n$ and $\alpha_n$:
\bee
h_n &=& h^{(0)}+c_1H_1(u^{(0)})^{-1/2}E_1^n+c_2H_2(u^{(0)})^{-1}E_2^n+
	\nonumber \\
&& + c_3H_3(u^{(0)})^{-3/2}E_3^n +c_1c_2H_4(u^{(0)})^{-3/2}E_1^n E_2^n+ 
	\nonumber\\
&& + c_1c_2^2H_5(u^{(0)})^{-5/2}E_1^n 
	E_2^{2n}+c_2^2H_6(u^{(0)})^{-2}E_2^{2n}+ \nonumber \\
&&+ c_1^2H_7(u^{(0)})^{-1}E_1^{2n}+c_1^2c_2H_8(u^{(0)})^{-2}E_1^{2n}E_2^n+
	\nonumber\\
&&+ c_1^2c_2^2H_9(u^{(0)})^{-3}E_1^{2n}E_2^{2n}, \\
\alpha_n &=& 
\alpha^{(0)} + 
c_1L_1(u^{(0)})^{-1/2}E_1^n+c_2L_2(u^{(0)})^{-1}E_2^n+\nonumber \\
&&+c_3L_3(u^{(0)})^{-3/2}E_3^n+c_1c_2L_4(u^{(0)})^{-3/2}E_1^nE_2^n+
	\nonumber \\
&&+c_1c_2^2L_5(u^{(0)})^{-5/2}E_1^nE_2^{2n}+c_2^2L_6(u^{(0)})^{-2}E_2^{2n}+ 
	\nonumber \\
&&+c_1^2L_7(u^{(0)})^{-1}E_1^{2n}+c_1^2c_2L_8(u^{(0)})^{-2}E_1^{2n}E_2^n+
	\nonumber \\
&&+c_1^2c_2^2L_9(u^{(0)})^{-3}E_1^{2n}E_2^{2n}, 
\nonumber 
\eee
where
\bee
&& H_1=\sqrt{6}-\frac{h^{(0)}w_{21}^{(0)}}{2},~~H_2=\sqrt{6}w_{12}^{(0)}-
	\frac{h^{(0)}}{2}, \nonumber \\
&& H_3=\sqrt{6}w_{13}^{(0)}-\frac{h^{(0)}w_{23}^
	{(0)}}{2}, \nonumber \\
&& H_4=\frac{3}{4}h^{(0)}w_{21}^{(0)}-\frac{\sqrt{6}}{2}(1+w_{12}^{(0)}
	w_{21}^{(0)}), \nonumber \\
&& H_5=\frac{3\sqrt{6}}{4}(\frac{1}{2}+w_{12}^{(0)}w_{21}^{(0)}-
	\frac{5}{4\sqrt{6}}h^{(0)}w_{21}^{(0)}), \nonumber \\
&& H_6=\frac{1}{2}(\frac{3}{4}h^{(0)}-\sqrt{6}w_{12}^{(0)}),~~
	H_7=\frac{w_{21}^{(0)}}{2}(\frac{3}{4}h^{(0)}w_{21}^{(0)}-\sqrt{6}),~~
	\nonumber \\
&& H_8=\frac{3\sqrt{6}}{4}w_{21}^{(0)}(1+\frac{1}{2}w_{12}^{(0)}w_{21}^{(0)}-
	\frac{5}{4\sqrt{6}}h^{(0)}w_{21}^{(0)}), \nonumber \\
&& H_9=\frac{15\sqrt{6}}{16}w_{21}^{(0)}(\frac{7}{4\sqrt{6}}h^{(0)}
	w_{21}^{(0)}-1-w_{12}^{(0)}w_{21}^{(0)}); \nonumber \\
&& L_1=\frac{\sqrt{6}}{15}w_{31}^{(0)}-\frac{3\alpha^{(0)}
	w_{21}^{(0)}}{2},~~L_2=\frac{\sqrt{6}}{15}w_{32}^{(0)}-
	\frac{3\alpha^{(0)}}{2}, \nonumber \\
&& L_3=\frac{\sqrt{6}}{15}-\frac{3\alpha^{(0)}w_{23}^{(0)}}{2},\\
&& L_4=\frac{15}{4}\alpha^{(0)}w_{21}^{(0)}-\frac{\sqrt{6}}{10}
	(w_{31}^{(0)}+w_{21}^{(0)}w_{32}^{(0)}), \nonumber \\
&& L_5=\frac{\sqrt{6}}{4}(\frac{1}{2}w_{31}^{(0)}+w_{21}^{(0)}
	w_{32}^{(0)}-\frac{105}{4\sqrt{6}}\alpha^{(0)}w_{21}^{(0)}), 
	\nonumber \\
&&L_6=\frac{1}{2}(\frac{15}{4}\alpha^{(0)}-\frac{\sqrt{6}}{5}
	w_{32}^{(0)}),\nonumber \\
&& L_7=\frac{w_{21}^{(0)}}{2}(\frac{15}{4}\alpha^{(0)}w_{21}^{(0)}-
	\frac{\sqrt{6}}{5}w_{31}^{(0)}), \nonumber \\
&& L_8=\frac{\sqrt{6}}{4}w_{21}^{(0)}(w_{31}^{(0)}+\frac{1}{2}w_{21}^{(0)}
	w_{32}^{(0)}-\frac{105}{4\sqrt{6}}\alpha^{(0)}w_{21}^{(0)}),
	\nonumber \\
&& L_9=\frac{7\sqrt{6}}{16}w_{21}^{(0)}(\frac{45}{4\sqrt{6}}\alpha^{(0)}
	w_{21}^{(0)}-w_{31}^{(0)}-w_{21}^{(0)}w_{32}^{(0)}).\nonumber 
\eee
Considering (2.4), we rewrite the partial energy of the $n$-th phase 
layer as
\bee
F_n &=& -kTN's^{-3n}[f_{CR}^{(0)}+f_{CR}^{(1)}(u^{(0)})^{-1/2}c_1E_1^n+
	f_{CR}^{(2)}(u^{(0)})^{-1}c_2E_2^n+ \nonumber\\
&&+ f_{CR}^{(3)}(u^{(0)})^{-3/2}c_3E_3^n+f_{CR}^{(4)}(u^{(0)})^{-3/2}
	c_1c_2E_1^nE_2^n+ \nonumber\\
&&+f_{CR}^{(5)}(u^{(0)})^{-5/2}c_1c_2^2E_1^nE_2^{2n}+ 
	f_{CR}^{(6)}(u^{(0)})^{-2}c_2^2E_2^{2n}+ \\
&&+f_{CR}^{(7)}(u^{(0)})^{-1}c_1^2E_1^{2n}+f_{CR}^{(8)}(u^{(0)})^{-2}c_1^2c_2E_1^{2n}E_2^n+ 
	\nonumber \\
&&+f_{CR}^{(9)}(u^{(0)})^{-3}c_1^2c_2^2E_1^{2n}E_2^{2n}]. \nonumber
\eee
Here,
\bee
f_{CR}^{(m)} &=& 
	H_m(\varphi_3+\varphi_1/E_m)+L_m(\varphi_4+\varphi_2/E_m), m=1,2,3,  
	\nonumber\\
f_{CR}^{(4)} &=& H_4(\varphi_3+\varphi_1/(E_1E_2))+L_4(\varphi_4+\varphi_2/
	(E_1E_2))+\nonumber \\
&&+2H_1H_2(\varphi_3^{'}+\varphi_1^{'}/(E_1E_2))
	+2L_1L_2(\varphi_4^{'}+\varphi_2^{'}/(E_1E_2))+\nonumber \\
&&+(H_1L_2+H_2L_1)(\varphi_6^{'}+\varphi_5^{'}/(E_1E_2)), \nonumber\\
f_{CR}^{(5)} &=& 
	H_5(\varphi_3+\varphi_1/(E_1E_2^2))+L_5(\varphi_4+\varphi_2/(E_1E_2^2))+
	\nonumber \\
&&+2(H_1H_6+H_2H_4)(\varphi_3^{'}+\varphi_1^{'}/(E_1E_2^2))+ 
	2(L_1L_6+L_2L_4)\times \nonumber\\
&&\times (\varphi_4^{'}+\varphi_2^{'}/(E_1E_2^2))+
	(H_1L_6+H_6L_1+H_2L_4+H_4L_2)\times \nonumber \\
&&\times (\varphi_6^{'}+\varphi_5^{'}/(E_1E_2^2)),\nonumber \\
f_{CR}^{(6)} &=& 
	H_6(\varphi_3+\varphi_1/E_2^2)+L_6(\varphi_4+\varphi_2/E_2^2)+H_2^2(\varphi_3^{'}+\varphi_1^{'}/E_2^2)+ 
	\nonumber \\
&&+L_2^2(\varphi_4^{'}+\varphi_2^{'}/E_2^2)+H_2L_2(\varphi_6^{'}+
	\varphi_5^{'}/E_2^2),	 \\
f_{CR}^{(7)} &=& H_7(\varphi_3+\varphi_1/E_1^2)+L_7(\varphi_4+
	\varphi_2/E_1^2)+H_1^2(\varphi_3^{'}+\varphi_1^{'}/E_1^2)+\nonumber \\
&&+L_1^2(\varphi_4^{'}+
	\varphi_2^{'}/E_1^2)+H_1L_1(\varphi_6^{'}+\varphi_5^{'}/E_1^2),\nonumber \\
f_{CR}^{(8)} &=& H_8(\varphi_3+\varphi_1/(E_1^2E_2))+L_8
	(\varphi_4+\varphi_2/(E_1^2E_2))+2(H_1H_4+\nonumber \\
&&+H_2H_7)(\varphi_3^{'}+\varphi_1^{'}/(E_1^2E_2))+2(L_1L_4+L_2L_7)(\varphi_4^{'}+\nonumber 
	\\
&&+\varphi_2^{'}/(E_1^2E_2))+(H_1L_4+H_4L_1+H_2L_7+H_7L_2)\times \nonumber\\
&&\times (\varphi_6^{'}+\varphi_5^{'}/(E_1^2E_2)), 
	\nonumber \\
f_{CR}^{(9)} &=& H_9(\varphi_3+\varphi_1/(E_1E_2)^2)+L_9
	(\varphi_4+\varphi_2/(E_1E_2)^2)
	+(2H_1H_5+\nonumber \\	
&&+2H_2H_8+H_4^2+2H_6H_7) 
	(\varphi_3^{'}+\varphi_1^{'}/(E_1E_2)^2)+(2L_1L_5+\nonumber\\
&&+2L_2L_8+L_4^2+2L_6L_7)(\varphi_4^{'}+\varphi_2^{'}/(E_1E_2)^2)+ 
	(H_1L_5+\nonumber \\
&&+H_5L_1+H_2L_8+H_8L_2+H_4L_4+H_6L_7+H_7L_6)\times \nonumber \\
&&\times (\varphi_6^{'}+\varphi_5^{'}/(E_1E_2)^2). 
\nonumber 
\eee
The quantities $u^{(0)}, w_{il}^{(0)}$ were determined in [13]. Let us note 
that in the expressions for $h_n$ and $\alpha_n$ we can neglect a 
qualitatively new term proportional to $E_3^n$ which arises within 
the $\rho^6$ model considered ($E_3$ is not essential as compared to $E_1$
or $E_2$).

Summing up expressions for $F_n$ (2.6) over layers of the phase 
space from $n=0$ to $n=m_{\tau}$, we obtain
\bee
F_{CR}&=&F_0^{'}+F_{CR}^{'}, \nonumber \\
F_0^{'}&=&-kTN^{'}[\ln Q({\cal M})+\ln Q(d)], \\
F_{CR}^{'}&=&F_{CR}^{(0)}+F_{CR}^{(1)}\tau^{3\nu}+F_{CR}^{(2)}\tau^{3\nu},
\nonumber 
\eee  
where
\bee
F_{CR}^{(0)}&=&-kTN's^{-3}[\frac{f_{CR}^{(0)}}{1-s^{-3}}+
	\frac{f_{CR}^{(1)}\varphi_0^{-1/2}\tilde{c}_1\tau E_1}{1-E_1s^{-3}}+ 
	\nonumber \\
&& +\frac{f_{CR}^{(2)}\varphi_0^{-1}c_{20}E_2}{1-E_2s^{-3}}+
	\frac{f_{CR}^{(3)}\varphi_0^{-3/2}c_{30}E_3}{1-E_3s^{-3}}+
	\frac{f_{CR}^{(4)}\varphi_0^{-3/2}\tilde{c}_1\tau c_{20}E_1E_2}{1-E_1E_2s^
	{-3}}+ \nonumber \\
&&+\frac{f_{CR}^{(5)}\varphi_0^{-5/2}\tilde{c}_1\tau c_{20}^2 E_1E_2^2}
	{1-E_1E_2^2s^{-3}}+\frac{f_{CR}^{(6)}\varphi_0^{-2} c_{20}^2 E_2^2}{1-E_2^2
	s^{-3}}+\nonumber \\
&&+\frac{f_{CR}^{(7)}\varphi_0^{-1}\tilde{c}_1^2\tau^2 
	E_1^2}{1-E_1^2s^{-3}}+\frac{f_{CR}^{(8)}\varphi_0^{-2}\tilde{c}_1^2\tau^2 
	c_{20}E_1^2E_2}{1-E_1^2E_2s^{-3}}+ \nonumber \\
&&+\frac{f_{CR}^{(9)}\varphi_0^{-3}\tilde{c}_1^2\tau^2 c_{20}^2E_1^2E_2^2}
	{1-E_1^2E_2^2s^{-3}}],\\
F_{CR}^{(1)}&=&kTN's^{-3m_0}[\frac{f_{CR}^{(0)}}{1-s^{-3}}+
	\frac{f_{CR}^{(1)}\varphi_0^{-1/2} f_0}{1-E_1s^{-3}}+ \nonumber \\
&&+\frac{f_{CR}^{(7)}\varphi_0^{-1}f_0^2}{1-E_1^2s^{-3}}],\nonumber \\
F_{CR}^{(2)}&=&-kTN's^{-3m_0}[\frac{f_{CR}^{(2)}\varphi_0^{-1}c_{20}E_2^{m_
	{\tau}+1}}{1-E_2s^{-3}}+
	\frac{ f_{CR}^{(3)} \varphi_0^{-3/2} c_{30} E_3^{m_{\tau}+1}} 
	{1-E_3s^{-3}}+ \nonumber \\
&&+\frac{f_{CR}^{(4)}\varphi_0^{-3/2}f_0c_{20}E_2^{m_{\tau}+1}}{1-
	E_1E_2s^{-3}}+\frac{f_{CR}^{(5)}\varphi_0^{-5/2}f_0c_{20}^2
	E_2^{2(m_{\tau}+1)}}{1-E_1E_2^2s^{-3}}+\nonumber \\
&&+\frac{f_{CR}^{(6)}\varphi_0^{-2}c_{20}^2E_2^{2(m_{\tau}+1)}}{1-E_2^2s^{-3}}+\frac{f_{CR}^{(8)}\varphi_0^{-2}f_0^2 
	c_{20}E_2^{m_{\tau}+1}}{1-E_1^2E_2s^{-3}}+\nonumber \\
&&+\frac{f_{CR}^{(9)}\varphi_0^{-3}f_0^2 c_{20}^2 
E_2^{2(m_{\tau}+1)}}{1-E_1^2E_2^2s^{-3}}].\nonumber
	\eee
To calculate these expressions, we have used the formulas:
\bee
s^{-3(m_{\tau}+1)}&=&\tau^{3\nu}s^{-3m_0}, 
\\
c_1=\tau\tilde{c}_1\beta\tilde{\Phi}(0),\quad 
c_2&=&c_{20}(\beta\tilde{\Phi}(0))^2, 
\quad c_3=c_{30}(\beta\tilde{\Phi}(0))^3, \nonumber
\eee
where $m_0, \tilde{c}_1, c_{20}, c_{30}$ and $f_0, \varphi_0$
were defined in [14]. The dependence of $m_0$ on s for the 
$\rho^6$ model under consideration is plotted in figure 1 (solid curve). 
Here the dashed curve corresponds to the $\rho^4$ model. It is easy to 
notice that, when  $s=s^{*}$ ($s^{*}$ is the value at which $h_n$ turns 
to zero at the fixed point; $s^{*}$=3.5862 for the $\rho^4$ model  and 
$s^{*}$=2.7349 for the $\rho^6$ model), the values of $m_0$ for $\rho^4$ and 
$\rho^6$ models coincide.
\begin{figure}[htb]
\epsfxsize 85mm
\epsfysize 45mm
\centerline{\epsffile{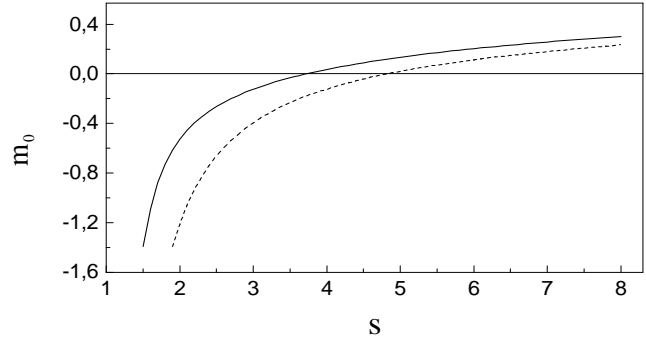}}
\caption {Dependence of the quantity $m_0$ on the RG parameter $s$ for the 
$\rho^6$ model (solid curve) and the $\rho^4$ model (dashed curve).}
\end{figure}
 
Further, we put $F_{CR}^{(2)} =0$ at $\tau \ll 1$, 
since $E_2 < 1, E_3 < 1$, and  $m_{\tau}$ is large. Taking $E_2, E_3$ into 
account gives rise to terms characterising cor\-rec\-tions to scaling. As a 
result, the free energy of the CR region takes the form:
\bee
F_{CR}&=&-kTN'[{\gamma_0+\delta_0-\gamma_3^{(CR)+}\tau^{3\nu}}], \nonumber\\
\gamma_0&=&s^{-3}[\frac{f_{CR}^{(0)}}{1-s^{-3}}+\frac{f_{CR}^{(1)}\varphi_0^{-1/2}
	\tilde{c}_1\tau E_1}{1-E_1s^{-3}}+\frac{f_{CR}^{(2)}\varphi_0^{-1/2}
	c_{20} E_2}{1-E_2s^{-3}}+ \nonumber \\
&&+\frac{f_{CR}^{(3)}\varphi_0^{-3/2}c_{30}E_3}{1-E_3s^{-3}}+
	\frac{f_{CR}^{(4)}\varphi_0^{-3/2}\tilde{c}_1\tau 
	c_{20}E_1E_2}{1-E_1E_2s^{-3}}+\nonumber \\
&&+\frac{f_{CR}^{(5)}\varphi_0^{-5/2}\tilde{c}_1c_{20}^2\tau E_1 
	E_2^2}{1-E_1E_2^2s^{-3}}+\frac{f_{CR}^{(6)}\varphi_0^{-2}c_{20}^2 
	E_2^2}{1-E_2^2s^{-3}}+
	\frac{f_{CR}^{(7)}\varphi_0^{-1}\tilde{c}_1^2{\tau}^2 
	E_1^2}{1-E_1^2s^{-3}}+ \nonumber\\
&&+\frac{f_{CR}^{(8)}\varphi_0^{-2}\tilde{c}_1^2c_{20}{\tau}^2 
	E_1^2E_2}{1-E_1^2E_2s^{-3}}+\frac{f_{CR}^{(9)}\varphi_0^{-3}\tilde{c}_1^2c_{20}^2{\tau}^2 
	E_1^2E_2^2}{1-E_1^2E_2^2s^{-3}}], \\	
\delta_0&=&\ln Q({\cal M})+\ln Q(d), \qquad 
\gamma_3^{(CR)+}=c_{\nu}^3\gamma^{+}, \qquad 
c_{\nu}=(\frac{\tilde{c}_1^{(0)}}{f_0})^{\nu},\nonumber \\
\gamma^{+}&=&\frac{f_{CR}^{(0)}}{1-s^{-3}}+\frac{f_{CR}^{(1)}\varphi_0^{-1/2}
	f_0}{1-E_1s^{-3}}+\frac{f_{CR}^{(7)}\varphi_0^{-1}f_0^2}{1-E_1^2s^{-3}}\nonumber.
\eee

Note that $\gamma_0, \delta_0$ are the functions of temperature, since 
they are expressed in terms of $\tilde{c}_1, c_{20}, c_{30}$ and 
$Q(d)$. Let us extract the temperature dependence in these quantities.

Near $T_c$ we have for $\tilde{c}_1$
\bee
\tilde{c}_1&=&\tilde{c}_1^{(0)}+\tilde{c}_1^{(1)}\tau,\nonumber\\
\tilde{c}_1^{(0)}&=&V_1[1-f_0+v_{12}^{(0)}\varphi_0^{1/2}+v_{13}^{(0)}\psi_0\varphi_0^{-1}+
	\frac{a_4^{'}v_{12}^{(0)}\varphi_0^{-1/2}}{(\beta_c 
	\tilde{\Phi}(0))^2}+ \nonumber \\
&+&\frac{2a_6^{'}v_{13}^{(0)}\varphi_0^{-1}}{(\beta_c 
	\tilde{\Phi}(0))^3}], \\
\tilde{c}_1^{(1)}&=&V_1[\frac{a_4^{'}v_{12}^{(0)}
\varphi_0^{-1/2}}{(\beta_c\tilde{\Phi}(0))^2}+\frac{3a_6^{'}v_{13}^{(0)}
\varphi_0^{-1}}{(\beta_c\tilde{\Phi}(0))^3}],\nonumber 
\eee 
for $c_{20}$ and $c_{30}$ we get, respectively, 
\bee
c_{20}&=&c_{20}^{(0)}+c_{20}^{(1)}\tau+c_{20}^{(2)}\tau^2, \nonumber\\
c_{20}^{(0)}&=&V_2[-\varphi_0-v_{21}^{(0)}(1-f_0)\varphi_0^{1/2}-
v_{23}^{(0)}\psi_0\varphi_0^{-1/2}+
	\frac{a_2^{'}v_{21}^{(0)}\varphi_0^{1/2}}{\beta_c\tilde{
	\Phi}(0)}+\nonumber \\
&+&\frac{a_4^{'}}{(\beta_c\tilde{\Phi}(0))^2}+ 
	\frac{a_6^{'}v_{23}^{(0)}\varphi_0^{-1/2}}{(\beta_c 
	\tilde{\Phi}(0))^3}], \\
c_{20}^{(1)}&=&V_2[\frac{a_2^{'}v_{21}^{(0)}\varphi_0^{1/2}}{\beta_c\tilde{
	\Phi}(0)}+\frac{2a_4^{'}}{(\beta_c\tilde{\Phi}(0))^2}+ 
	\frac{3a_6^{'}v_{23}^{(0)}\varphi_0^{-1/2}}{(\beta_c 
	\tilde{\Phi}(0))^3}], \nonumber \\
c_{20}^{(2)}&=&V_2[\frac{a_4^{'}}{(\beta_c\tilde{\Phi}(0))^2}+ 
	\frac{3a_6^{'}v_{23}^{(0)}\varphi_0^{-1/2}}{(\beta_c 
	\tilde{\Phi}(0))^3}]; \nonumber
\eee 
\bee
c_{30}&=&c_{30}^{(0)}+c_{30}^{(1)}\tau+c_{30}^{(2)}\tau^2, \nonumber\\
c_{30}^{(0)}&=&V_3[-\psi_0-\varphi_0^{3/2}v_{32}^{(0)}-\varphi_0(1-f_0)
	v_{31}^{(0)}+\frac{a_2^{'}v_{31}^{(0)}\varphi_0} {\beta_c\tilde{\Phi}(0)}+ 
	\nonumber \\
&+&\frac{a_4^{'}v_{32}^{(0)}\varphi_0^{1/2}}{(\beta_c\tilde{\Phi}(0))^2}+ 
	\frac{a_6^{'}}{(\beta_c\tilde{\Phi}(0))^3}], \\
c_{30}^{(1)}&=&V_3[\frac{a_2^{'}v_{31}^{(0)}\varphi_0}{\beta_c\tilde{
	\Phi}(0)}+\frac{2a_4^{'}v_{32}^{(0)}\varphi_0^{1/3}}
	{(\beta_c\tilde{\Phi}(0))^2}+ 
	\frac{3a_6^{'}}{(\beta_c\tilde{\Phi}(0))^3}], \nonumber\\
c_{30}^{(2)}&=&V_3[\frac{a_4^{'}v_{32}^{(0)}\varphi_{0}^{1/2}}
	{(\beta_c\tilde{\Phi}(0))^2}+\frac{3a_6^{'}}{(\beta_c\tilde{\Phi}(0))^3}]. 
\nonumber
\eee
The values of $\beta_c\tilde{\Phi}(0)$, the correlation length critical 
exponent $\nu=\ln s/ \ln E_1$, exponents of the corrections to 
scaling $\Delta_1 = -\ln~ E_2~/ \ln~ E_1$, ~$\Delta_2 = -\ln E_3~/ \ln E_1$, 
and coefficients of the expressions for $\tilde{c}_1, c_{20}, c_{30}$
are given in tables 1,2. In the present paper, 
the numerical calculations are performed at $b/c=1$ and arithmetically 
averaged Fourier transform of the potential.

\begin{table}[htb]
\caption{Values of $\beta_c\tilde{\Phi}(0), \nu, \Delta_1, \Delta_2,
\tilde{c}_1^{(0)}, \tilde{c}_1^{(1)}$ for different $s$.}
\vspace{1em}
\begin{tabular}{lllllll}
\hline	
 &  &  &  &  &  & \\
~$s$~&$\beta_c\tilde{\Phi}(0)$~&~$\nu
$~&~$\Delta_1$~&$\Delta_2$~&~$\tilde c_1^{(0)}$~&~$\tilde c_1^{(1)}$\\
 &  &  &  &  &  & \\
\hline
 &  &  &  &  &  & \\
~~2~~~ &~~ 1.1204~~~& 0.619~~~ & 0.653~~~&  5.061~~~ & 0.7602~~~& 0.0109\\
~~2.5    &~~ 1.1405 & 0.634 &  0.552 &  3.963 & 0.7637 & 0.0103 \\
~~2.7349 &~~ 1.1506 & 0.637 &  0.525 &  3.647 & 0.7641 & 0.0100 \\
~~3      &~~ 1.1628 & 0.640 &  0.503 &  3.379 & 0.7630 & 0.0096 \\
~~3.5    &~~ 1.1882 & 0.645 &  0.476 &  3.038 & 0.7570 & 0.0089 \\
~~3.5862 &~~ 1.1929 & 0.645 &  0.473 &  2.994 & 0.7555 & 0.0088 \\
~~4      &~~ 1.2165 & 0.648 &  0.460 &  2.821 & 0.7471 & 0.0082 \\
 &  &  &  &  &  & \\
\hline
\end{tabular}
\end{table}

\begin{table}[htb]
\caption{Coefficients in equations for $c_{20}$ (2.13) and 
$c_{30}$ (2.14).}
\vspace{0.3mm}
\begin{tabular}{lllllll}
&  &  &  &  &  & \\
\hline
&  &  &  &  &  & \\
~~$s$~~&$c_{20}^{(0)}$~&~$c_{20}^{(1)}
$~~&~$c_{20}^{(2)}$~~&$c_{30}^{(0)}$~&~~$c_{30}^{(1)}$~~&~~
$c_{30}^{(2)}$\\
&  &  &  &  &  & \\
\hline
&  &  &  &  &  & \\
2~~~~& -0.2302~~~& -0.0949~~~ & 0.0075 
~~~&  0.2401~~~ & 0.0762~~~& -0.0161 \\
2.5    & -0.3297 & -0.0893 & 0.0080  & 0.3266  & 0.0698 & -0.0147 \\
2.7349 & -0.3820 & -0.0872 &  0.0083 &  0.3642 & 0.0654 & -0.0140 \\
3      & -0.4449 & -0.0851 &  0.0085 &  0.4056 & 0.0603 & -0.0133 \\
3.5    & -0.5737 & -0.0813 &  0.0089 &  0.4821 & 0.0516 & -0.0120 \\
3.5862 & -0.5972 & -0.0807 &  0.0089 &  0.4952 & 0.0503 & -0.0118 \\
4      & -0.7142 & -0.0776 &  0.0090 &  0.5581 & 0.0443 & -0.0108 \\
 &  &  &  &  &  & \\
\hline
\end{tabular}
\end{table}

Expressions for the quantities $V_1, V_2, V_3,\psi_0, v_{l j}^{(0)}$ 
occurring in (2.12) - (2.14) are presented in [13,14]. The coefficient 
$\gamma_0$ can be written as
\bee
\gamma_0&=&\gamma_0^{(0)}+\gamma_0^{(1)}\tau+\gamma_0^{(2)}\tau^2, 
\nonumber \\
\gamma_0^{(0)}&=&s^{-3}[\frac{f_{CR}^{(0)}}{1-s^{-3}}+
	\frac{f_{CR}^{(2)}\varphi_0^{-1}c_{20}^{(0)}E_2}{1-E_2 s^{-3}}+ 
	\frac{f_{CR}^{(3)}\varphi_0^{-3/2}c_{30}^{(0)}E_3}{1-E_3 
	s^{-3}}+\nonumber \\
&+&\frac{f_{CR}^{(6)}\varphi_0^{-2}(c_{20}^{(0)})^2 E_2^2}{1-E_2^2 
	s^{-3}}],\nonumber \\
\gamma_0^{(1)}&=&s^{-3}[\frac{f_{CR}^{(1)}\varphi_0^{-1/2}\tilde{c}_1^{(0)}E_1}{1-E_1s^{-3}}+
	\frac{f_{CR}^{(2)}\varphi_0^{-1}c_{20}^{(1)}E_2}{1-E_2 s^{-3}}+ 
	\frac{f_{CR}^{(3)}\varphi_0^{-3/2}c_{30}^{(1)}E_3}{1-E_3 s^{-3}}+\nonumber 
\\
&+&\frac{f_{CR}^{(4)}\varphi_0^{-3/2}\tilde{c}_1^{(0)}c_{20}^{(0)}E_1
	E_2}{1-E_1E_2 
	s^{-3}}+\frac{f_{CR}^{(5)}\varphi_0^{-5/2}\tilde{c}_1^{(0)}(c_{20}^{(0)})^2
	E_1E_2^2}{1-E_1E_2^2s^{-3}}+\nonumber \\
&+&\frac{2f_{CR}^{(6)}\varphi_0^{-2}c_{20}^{(0)}c_{20}^{(1)} E_2^2}{1-E_2^2 
	s^{-3}}],\\
\gamma_0^{(2)}&=&s^{-3}[\frac{f_{CR}^{(1)}\varphi_0^{-1/2}\tilde{c}_1^{(1)}E_1}
	{1-E_1s^{-3}}+
	\frac{f_{CR}^{(2)}\varphi_0^{-1}c_{20}^{(2)}E_2}{1-E_2 s^{-3}}+ 
	\frac{f_{CR}^{(3)}\varphi_0^{-3/2}c_{30}^{(2)}E_3}{1-E_3 s^{-3}}+\nonumber 
\\
&+&\frac{f_{CR}^{(4)}\varphi_0^{-3/2}(\tilde{c}_1^{(0)}c_{20}^{(1)}+\tilde{c}_1
	^{(1)}c_{20}^{(0)})E_1E_2}{1-E_1E_2s^{-3}}+\nonumber \\
&+&\frac{f_{CR}^{(5)}\varphi_0^{-5/2}(2\tilde{c}_1^{(0)}c_{20}^{(0)}c_{20}^{(1)}+
	\tilde{c}_1^{(1)}(c_{20}^{(0)})^2)E_1E_2^2}
	{1-E_1E_2^2s^{-3}}+\nonumber \\
&+&\frac{f_{CR}^{(6)}\varphi_0^{-2}((c_{20}^{(1)})^2+2c_{20}^{(0)}c_{20}^{(2)})E_2^2}
	{1-E_2^2 s^{-3}}+
	\frac{f_{CR}^{(7)}\varphi_0^{-1}(\tilde{c}_{1}^{(0)})^2 E_1^2}
	{1-E_1^2 s^{-3}}+ \nonumber \\
&+&\frac{f_{CR}^{(8)}\varphi_0^{-2}(\tilde{c}_{1}^{(0)})^2c_{20}^{(0)}E_1^2E_2}
	{1-E_1^2 E_2 s^{-3}}
	+\frac{f_{CR}^{(9)}\varphi_0^{-3}(\tilde{c}_{1}^{(0)})^2(c_{20}^{(0)})^2 
	E_1^2E_2^2}{1-E_1^2E_2^2 s^{-3}}]. \nonumber 
\eee
For $\delta_0$ we obtain
\bee
\delta_0&=&\delta_0^{(0)}+\delta_0^{(1)}\tau+\delta_0^{(2)}\tau^2, 
	\nonumber \\
\delta_0^{(0)}&=&\ln Q({\cal M}) + \ln Q(d,T_c), \nonumber \\
\delta_0^{(1)}&=&-\frac{\sqrt{6}}{\sqrt{a_4^{'}}}(1-\bar{q})\beta_c 
	\tilde{\Phi}(0) {\cal F}_2(h_c,\alpha), \\
\delta_0^{(2)}&=&-\frac{3}{a_4^{'}}(1-\bar{q})^2 (\beta_c 
	\tilde{\Phi}(0))^2 [{\cal F}_2^2(h_c,\alpha) - {\cal F}_4(h_c,\alpha)] +
	\nonumber \\
&+&\frac{\sqrt{6}}{\sqrt{a_4^{'}}}(1-\bar{q})\beta_c 
\tilde{\Phi}(0) {\cal F}_2(h_c,\alpha), \nonumber \\
h_c&=&\frac{\sqrt{6}}{\sqrt{a_4^{'}}}[a_2^{'}-\beta_c\tilde{\Phi}(0)
	(1-\bar{q})], \quad 
	\alpha = \frac{\sqrt{6}}{15}\frac{a_6^{'}}{(a_4^{'})^{3/2}}, \quad 
	\bar{q}=\frac{1+s^{-2}}{2}. \nonumber
\eee
Coefficients of the expressions for $\gamma_0$ (2.15), $\delta_0$ (2.16) 
at different values of the RG parameter $s$ are given in table 3.
\begin{table}[htb]
\caption{Values of coefficients in equations for $\gamma_0$ 
(2.15) and $\delta_0$ (2.16).}
\vspace{0.3mm}
\begin{tabular}{lllllll}
&  &  &  &  &  & \\
\hline
&  &  &  &  &  & \\
~~$s$~~&$\gamma_0^{(0)}$~&~$\gamma_0^{(1)}
$~~&~$\gamma_0^{(2)}$~~&$\delta_0^{(0)}$~&~~$\delta_0^{(1)}$~~&~~
$\delta_0^{(2)}$\\
&  &  &  &  &  & \\
\hline
&  &  &  &  &  & \\
2~~~ & 0.1205~~~& -0.3160~~~ & -2.5291 
~~~&  0.2711~~~ & -0.3574~~~& 0.4810 \\
2.5    & 0.0757 & -0.2343 &  -2.9320 &  0.3237 & -0.4510 & 0.6464 \\
2.7349 & 0.0624 & -0.2062 &  -2.9801 &  0.3423 & -0.4862 & 0.7128 \\
3      & 0.0510 & -0.1798 &  -2.9798 &  0.3606 & -0.5222 & 0.7827 \\
3.5    & 0.0362 & -0.1415 &  -2.9314 &  0.3905 & -0.5836 & 0.9075 \\
3.5862 & 0.0343 & -0.1360 &  -2.9236 &  0.3953 & -0.5938 & 0.9287 \\
4      & 0.0267 & -0.1135 &  -2.9039 &  0.4175 & -0.6420 & 1.0319 \\
&  &  &  &  &  & \\
\hline
\end{tabular}
\end{table}

Hence, the free energy of the CR region reads
\bee
F_{CR}&=&-kTN^{'}[\gamma_0^{(CR)}+\gamma_1\tau+\gamma_2\tau^2+
\gamma_{3}^{(CR)+}\tau^{3\nu}], \nonumber \\ 
\gamma_0^{(CR)}&=&\gamma_0^{(0)}+\delta_0^{(0)}, \\
\gamma_1&=&\gamma_0^{(1)}+\delta_0^{(1)}, \nonumber \\
\gamma_2&=&
\gamma_0^{(2)}+\delta_0^{(2)}. \nonumber
\eee  
The numerical values of coefficients $\gamma_0^{(CR)}$, $\gamma_1, 
\gamma_2$ and quantities $\gamma^{+}, 
c_{\nu}=(\tilde{c}_1^{(0)}/f_0)^{\nu}$ occurring in $\gamma_3^{(CR)+}$
are given in table 4.
\begin{table}[htb]
\caption{Coefficients $\gamma_0^{(CR)}$, $\gamma_1$, 
$\gamma_2$ and quantities $c_{\nu}, \gamma^{+}, \gamma^{-}$ contained in 
$\gamma_3^{(CR)\pm} $ for different values of the parameter $s$.}
\vspace{0.3mm}
\begin{tabular}{llllllr}
&  &  &  &  &  & \\
\hline
&  &  &  &  &  & \\
~~$s$~~&$\gamma_0^{(CR)}$~&~$\gamma_1
$~~&~$\gamma_2$~~&$c_{\nu}$~&~~$\gamma^+$~~&~~
$\gamma^-$\\
&  &  &  &  &  & \\
\hline
&  &  &  &  &  & \\
2~~~ & 0.3915~~~& -0.6734~~~ & -2.0482 
~~~&  1.4412~~~& -0.3170~~~& 0.7382 \\
2.5    & 0.3994 & -0.6852 &  -2.2856 &  1.2722 & -0.7757 & 0.5244 \\
2.7349 & 0.4047 & -0.6924 &  -2.2672 &  1.2097 & -0.9831 & 0.4188 \\ 
3      & 0.4116 & -0.7020 &  -2.1971 &  1.1462 & -1.2229 & 0.2899 \\
3.5    & 0.4267 & -0.7251 &  -2.0239 &  1.0414 & -1.7450 & -0.0277 \\
3.5862 & 0.4295 & -0.7298 &  -1.9949 &  1.0250 & -1.8496 & -0.0973 \\
4      & 0.4442 & -0.7555 &  -1.8720 &  0.9517 & -2.4345 & -0.5148 \\
&  &  &  &  &  & \\
\hline
\end{tabular}
\end{table}

Knowledge of $F_{CR}$ allows one to calculate other thermodynamic 
functions of the system in the CR region at $T>T_c$. For the entropy 
$S_{CR}$, internal energy $U_{CR}$ and specific heat $C_{CR}$ we get
\bee
S_{CR}&=&kN^{'}[s^{(0)(CR)}+c_0\tau +u_3^{(CR)+}\tau^{1-\alpha}],\nonumber\\
U_{CR}&=&kTN^{'}[\gamma_{1}+u_{1}\tau+u_3^{(CR)+}\tau^{1-\alpha}], 
\nonumber\\
C_{CR}&=&kN^{'}[c_0+c_3^{(CR)+}\tau^{-\alpha}], \\
s^{(0)(CR)}&=&\gamma_0^{(CR)}+\gamma_{1}, \quad 
c_{0}=2(\gamma_1+\gamma_2),\quad u_3^{(CR)+}=-3\nu \gamma_3^{(CR)+}, 
\nonumber \\
u_1&=&2\gamma_2+\gamma_1, \quad c_3^{(CR)+}=-3\nu (3\nu 
-1)\gamma_3^{(CR)+},\quad \alpha=2-3\nu. \nonumber
\eee

In the region of LGR, the expression for the part 
$Z_{LGR}$ of the partition function  (2.1) reads

\bee
Z_{LGR}&=&\int \exp \{-\frac{1}{2}\sum_{k \leq 
	B_{m_{\tau}+1}}[d_{m_{\tau}}(k)-d_{m_{\tau}}(B_{m_{\tau+1}},B_{m_{\tau}})]\rho_{\vec{k}}\rho_{-\vec{k}}- 
	\nonumber \\
&-&2\pi i \sum_{k \leq 
	B_{m_{\tau}+1}}\omega_{\vec{k}}\rho_{\vec{k}}-\frac{(2\pi)^2}{2}\sum_{k 
	\leq B_{m_{\tau}+1}}
	P_2^{(m_{\tau})}\omega_{\vec{k}}\omega_{-\vec{k}}- \\
&-&\sum_{l=2}^3 
	\frac{(2\pi)^{2l}}{(2l)!}{N_{m_{\tau}+1}^{1-l}}\sum_{k_1,\ldots,k_{2l}\leq 
	B_{m_{\tau}+1}}P_{2l}^{(m_{\tau})}\omega_{\vec{k}_1}\ldots 
	\omega_{\vec{k}_{2l}}\times \nonumber\\
&\times&\delta_{\vec{k}_1+ \cdots 
	+\vec{k}_{2l}}\}(d\rho)^{N_{m_{\tau}+1}}(d \omega)^{N_{m_{\tau}+1}}. 
	\nonumber
\eee 
To calculate $Z_{LGR}$ it is convenient to select two regions of the 
wave vector values [14]. The first, transition region, corresponds to the 
values of $k$ close to $B_{m_{\tau}}$; the second, Gaussian region, 
corresponds to  small values of the wave vector $(k \rightarrow 0)$.
Hence, we have
\bee
Z_{LGR}= Z_{LGR}^{(1)}Z_{LGR}^{(2)}.
\eee 

The contribution to free energy from the phase space layers following 
the point of exit from the CR region is
\bee
&&F_{LGR}^{(1)}=-kTN^{'}f_{TR}\tau^{3\nu}, \nonumber \\
&&f_{TR}=c_{\nu}^3\bar{f}_{TR}, \quad 
	\bar{f}_{TR}=\sum_{m=0}^{\tilde{m}_0}s^{-3m}f_{LGR_1}(m), 
\\&&f_{LGR_1}(m)=\ln \frac{2}{\pi}+\frac{1}{4}\ln 24 - \frac{1}{4}\ln 
	C(\eta_{m_{\tau}+m},\xi_{m_{\tau}+m})+ \nonumber \\
&&+ \ln I_0(h_{m_{\tau}+m+1},\alpha_{m_{\tau}+m+1})+\ln 
	I_0(\eta_{{m_{\tau}}+m},\xi_{m_{\tau}+m}), \nonumber 
\eee
where $\tilde{m}_0$ is the nearest integer to ${\tilde{m}_0}^{'}$.
The quantities ${\tilde{m}_0}^{'}$, $\eta_{m_{\tau}+m}$, $\xi_{m_{\tau}+m}$,
$h_{m_{\tau}+m+1}$, $\alpha_{m_{\tau}+m+1}$ were determined in 
[14]. The plots of ${\tilde{m}_0}^{'}(s)$ and analogous dependence 
${m_0}^{''}(s)$ ($\rho^4$ model) are represented in figure 2. 
\begin{figure}[htb]
\epsfxsize 85mm
\epsfysize 45mm
\centerline{\epsffile{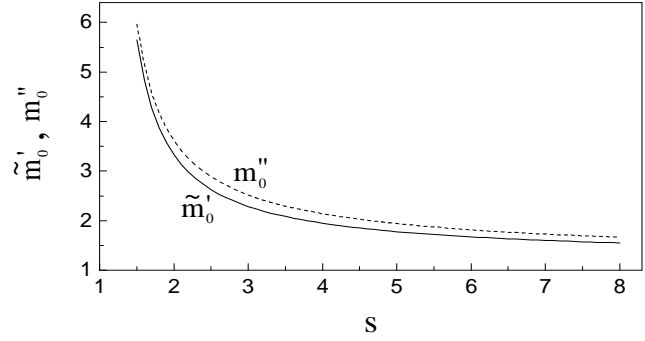}}
\caption {Behaviour of ${\tilde{m}_0}^{'}$ ($\rho^6$ model) and 
$m_0^{''}$ ($\rho^4$ model) with the increase of the parameter $s$.}
\end{figure}

Introducing an infinitely small external magnetic field $h=\mu_B 
\cal{H}$ ($\mu_B$ is the Bohr magneton), we can write the part of the free 
energy corresponding to $Z_{LGR}^{(2)}$ as
\bee 
F_{LGR}^{(2)}&=&-kTN^{'}f^{'}\tau^{3\nu}-\beta N 
	\gamma_4^{+}h^2\tau^{-2\nu},~~f^{'}=c_{\nu}^3\bar{f}^{'}, \nonumber\\
	\bar{f}^{'}&=&s^{-3(\tilde{m}_0+1)}f_{LGR_2},\nonumber \\ 
f_{{LGR}_2}&=&-\frac{1}{4}\ln 24+\frac{1}{3}+\frac{1}{4}\ln 
	\bar{u}_{{m_{\tau}}^{'}-1}-\frac{1}{2}\ln (\tilde{G}+\frac{1}{s^2})- \\
&-&\frac{1}{2}\ln{\cal 
	F}_2(h_{{m_{\tau}}^{'}-1},\alpha_{{m_{\tau}}^{'}-1})-\tilde{G}s^{2}+(\tilde{G}s^2)
	^{3/2}\arctan[{(\tilde{G}s^2)^{-1/2}}],
	\nonumber \\
\gamma_4^{+}&=&c_{\nu}^{-2}\bar{\gamma}_4^{+}/(\beta\tilde{\Phi}(0)), 
	\quad \bar{\gamma}_4^{+}=s^{2\tilde{m}_0}/(2\tilde{G}).\nonumber
\eee
Here,
\bee
&&m_{\tau}^{'}=m_{\tau}+\tilde{m}_0+2, \quad {\bar u}_{m_{\tau}^{'}-1}=
u_{m_{\tau}^{'}-1}(\beta\tilde{\Phi}(0))^{-2}, \nonumber \\
&&\tilde{G}=({\bar u}_
{m_{\tau}^{'}-1}/24)^{1/2}[{\cal F}_2(h_{{m_{\tau}}^{'}-1},\alpha_
{{m_{\tau}}^{'}-1})]^{-1}-{\bar q}. \nonumber
\eee

A general expression describing the contribution of long-wave fluctuations 
to the free energy (LGR region) reads
\bee
F_{LGR}&=&-kTN^{'}f_{LGR}\tau^{3\nu}-\beta N{\gamma_4}^{+}h^2\tau^{-2\nu},\\
f_{LGR}&=&c_{\nu}^{3}\bar{f}_{LGR},~~~\bar{f}_{LGR}=\bar{f}_{TR}+\bar{f}^{'}.
\nonumber 
\eee 
The values of $\bar{f}_{TR},~~\bar{f}^{'},~~\bar{\gamma}_4^{+}$	are given
in table 5.
\begin{table}[htb]
\caption{Values of $\bar{f}_{TR}$, $\bar{f}^{'}$, $\bar{\gamma}_4^{+}$.}
\begin{center}
\begin{tabular}{llll}
~~~~~~&  &  & \\
\hline
~~~~~~&  &  & \\
~~~$s$~~&${\bar f}_{TR}$~&~$\bar f'\times 10^5
$~~&~$\bar{\gamma}_4^+$\\
~~~~~~&  &  & \\
\hline
&  &  & \\
~~~2~~~~~~ & 0.6529~~~~~~& 0.4749~~~~~~ & 2.7055 \\
~~~2.5    & 0.8142 & 0.0102 & 3.0870  \\
~~~2.7349 & 0.8824 & 0.2155 &  2.1737  \\ 
~~~3      & 0.9541 & 0.0659 &  2.1841  \\
~~~3.5    & 1.0756 & 0.0091 &  2.2058 \\
~~~3.5862 & 1.0950 & 0.0067 &  2.2098  \\
~~~4      & 1.1822 & 0.0016 &  2.2307 \\
~~~~~~&  &  & \\
\hline
\end{tabular}
\end{center}
\end{table}

The entropy, internal energy, specific heat corresponding to the LGR region 
are defined by the relations

\bee
&&S_{LGR}=kN^{'}u_3^{(LGR)}\tau^{1-\alpha}, \nonumber \\
&&U_{LGR}=kTN^{'}u_3^{(LGR)}\tau^{1-\alpha}, \\ 
&&C_{LGR}= kN^{'}c_3^{LGR}\tau^{-\alpha}, \nonumber \\
&&u_3^{(LGR)}=3\nu f_{LGR}, \nonumber \\
&&c_3^{(LGR)}=3\nu(3\nu -1)f_{LGR}.\nonumber 
\eee

\sect{Contributions to the thermodynamic func\-ti\-ons of the model 
from the critical and inverse Gaussian regime (CR and IGR) regions below 
$T_c$}

The free energy at $T<T_c$ can be written as [6,10] 
\be
F=F_0+F_{CR}+F_{IGR},
\ee
where $F_0=-kTN\ln 2$ is the free energy of the system of $N$ 
non-interacting spins, $F_{CR}$ is the contribution to the free energy 
from the short-wave fluctuation phases of the spin moment density 
(CR region), and $F_{IGR}$ is the contribution 
from the long-wave phases of the fluctuations (IGR region).

The number $\mu_{\tau}$	of the CV phase space layer, 
separating the short-wave and long-wave phases of fluctuations, is an 
important characteristic of the system. It is determined from the equation
\be
\frac{r_{\mu_{\tau}+1}-r^{(0)}}{r^{(0)}}=\delta.
\ee 
Here $r^{(0)}$ corresponds to the fixed point of the RR, 
$r_{\mu_{\tau}+1}$ is determined from the solutions of the RR equations 
(see, for example, [13,14]), $\delta$ is a constant $(\delta \leq 1)$. 
In the present paper we put $\delta=1$ (see [8]). Let us write an 
equation for $\mu_{\tau}$
\be
\mid \tau \mid \tilde{c}_1 E_1^{\mu_{\tau}+1}=f_0,
\ee 
the solution of which is
\bee
\mu_{\tau}=-\frac{\ln \mid \tau \mid}{\ln E_1}+\mu_0-1,
&& \mu_0=\frac{\ln f_0 - \ln \tilde{c}_1^{(0)}}{\ln E_1}.
\eee

We need to sum the partial free energies over the layers 
of the CV phase space to calculate $F_{CR}$. Extracting an explicit 
dependence on the layer number, using relations (3.3) and
\be
s^{-3(\mu_{\tau}+1)}={\mid \tau \mid}^{3\nu} s^{-3\mu_0},~~
s^{-3\mu_0 }= c_{\nu}^3, 
\ee
we obtain
\be
F_{CR}=-kTN^{'}[\gamma_0^{(CR)}-\gamma_1 \mid \tau \mid +\gamma_2 {\mid \tau 
\mid}^2 - \gamma_3^{(CR)-} {\mid \tau \mid}^{3\nu}].
\ee

Coefficients $\gamma_0^{(CR)}, \gamma_1, \gamma_2$ are determined in 
(2.17),
\bee
\gamma_3^{(CR)-}&=&c_{\nu}^3 \gamma^{-}, \\
\gamma^{-}&=& \frac{f_{CR}^{(0)}}{1-s^{-3}}-\frac{f_{CR}^{(1)}
	\varphi_{0}^{-1/2}f_0}{1-E_1s^{-3}}+\frac{f_{CR}^{(7)}
	\varphi_{0}^{-1}f_0^2}{1-E_1^2s^{-3}}.\nonumber
\eee 
The value of $\gamma^{-}$ is given in table 4.

The entropy, internal energy and specific heat of the system 
correspon\-ding to the CR region read
\bee
&&S_{CR}=kN^{'}[s^{(0)(CR)}-c_0\mid \tau \mid + 
	u_3^{(CR)-}{\mid \tau \mid}^{1-\alpha}], \nonumber \\
&&U_{CR}=kTN^{'}[\gamma_1 - u_1 \mid \tau \mid + 
u_3^{(CR)-}{\mid\tau\mid}^{1-\alpha}], \\ 
&&C_{CR}=kN^{'}[c_0 - c_3^{(CR)-}{\mid \tau 
	\mid}^{-\alpha}], \nonumber \\
&&u_3^{(CR)-}=3\nu \gamma_3^{(CR)-}, \nonumber \\
&&c_3^{(CR)-}=3\nu(3\nu-1)\gamma_3^{(CR)-}.\nonumber 
\eee

Let us calculate now the contribution to the free energy from the 
IGR region
\be
F_{IGR}=-kTN^{'}s^{-3(\mu_{\tau}+1)}\ln [\sqrt{2} 
Q(P_{\mu_{\tau}})] - kT\ln Z_{\mu_{\tau}+1},
\ee 
where
\bee
&&\sqrt{2}Q(P_{\mu_{\tau}})=(\frac{4 s^3 
	a_4^{(\mu_{\tau})}}{\pi^4C(h_{\mu_{\tau}},\alpha_{\mu_{\tau}})})^{1/4}
	I_0(\eta_{\mu_{\tau}},\xi_{\mu_{\tau}}), \nonumber \\
&&Z_{\mu_{\tau}+1}=\int \exp \{ -\frac{1}{2}\sum_{k \leq B_{\mu_{\tau}+1}}
	d_{\mu_{\tau}+1}(k)\rho_{\vec{k}}\rho_{-\vec{k}} - \nonumber \\
&&-\sum_{l=2}^3 \frac{a_{2l}^{(\mu_{\tau}+1)}}{(2l)!}N_{\mu_{\tau}+1}^{1-l}
	\sum_{k_1,\ldots,k_{2l}\leq B_{\mu_{\tau}+1}}\rho_{\vec{k}_1}\cdots 
	\rho_{\vec{k}_{2l}} \delta_{\vec{k}_1+ \cdots + 
	\vec{k}_{2l}}](d\rho)^{N_{\mu_{\tau}+1}}. \nonumber
\eee

Consider the first term on the RHS of (3.9). Making use of the relations 
\bee
r_{\mu_{\tau}}&=&-\bar{r}_{\mu_{\tau}}\beta \tilde{\Phi}(0), ~~
~~\bar{r}_{\mu_{\tau}}=f_0(1+E_1^{-1}), \nonumber \\
u_{\mu_{\tau}}&=&\bar{u}_{\mu_{\tau}} (\beta \tilde{\Phi}(0))^2, 
~~\bar{u}_{\mu_{\tau}}=\varphi_0-f_0\varphi_0^{1/2} w_{21}^{(0)} E_1^{-1}, 
\\
w_{\mu_{\tau}}&=&\bar{w}_{\mu_{\tau}} (\beta \tilde{\Phi}(0))^3, 
~~\bar{w}_{\mu_{\tau}}=\psi_0-f_0\varphi_0 w_{31}^{(0)}E_1^{-1}, 
\nonumber
\eee
we find 
\bee
&&h_{\mu_{\tau}}=\sqrt{6}\frac{\bar{q}-\bar{r}_{\mu_{\tau}}}{\sqrt{\bar{u}_
{\mu_{\tau}}}},\quad \alpha_{\mu_{\tau}}=\frac{\sqrt{6}}{15}\frac{\bar{w}_
{\mu_{\tau}}}{(\bar{u}_{\mu_{\tau}})^{3/2}}, \nonumber \\
&&\eta_{\mu_{\tau}}=\sqrt{6}s^{3/2}{\cal F}_2(h_{\mu_{\tau}},\alpha_
{\mu_{\tau}})[C(h_{\mu_{\tau}},\alpha_{\mu_{\tau}})]^{-1/2}, \quad \\
&&\xi_{\mu_{\tau}}=\frac{\sqrt{6}}{15}s^{-3/2}N(h_{\mu_{\tau}},\alpha_
{\mu_{\tau}})[C(h_{\mu_{\tau}},\alpha_{\mu_{\tau}})]^{-3/2}.\nonumber
\eee
The first term on the RHS of (3.9) is equal to
\bee
s^{-3(\mu_{\tau}+1)}\ln[\sqrt{2}Q(P_{\mu_{\tau}})]=\gamma_{g}{\mid\tau\mid}
	^{3\nu}+s^{-3(\mu_{\tau}+1)}\ln \frac{\sqrt{\beta 
	\tilde{\Phi}(0)}}{s^{\mu_{\tau}+1}}, \\
\gamma_{g}=c_{\nu}^3\bar{\gamma}_{g},~~\bar{\gamma}_{g}=\ln 
	[(\frac{4s^7\bar{u}_{\mu_{\tau}}}{\pi^4C(h_{\mu_{\tau}},\alpha_
	{\mu_{\tau}})})^{1/4}I_0(\eta_{\mu_{\tau}},\xi_{\mu_{\tau}})]. \nonumber
\eee

To find the second term on the RHS of (3.9), we need to calculate 
$Z_{\mu_{\tau}+1}$. The coefficients occurring in it  equal
\bee
&&d_{\mu_{\tau}+1}(k)=r_{\mu_{\tau}+1}s^{-2(\mu_{\tau}+1)}+\tilde{q}k^2,
	\quad \tilde{q}=2\beta\tilde{\Phi}(0)b^2, \nonumber \\
&&a_4^{(\mu_{\tau}+1)}=u_{\mu_{\tau}+1}s^{-4(\mu_{\tau}+1)}, \quad
	a_6^{(\mu_{\tau}+1)}=w_{\mu_{\tau}+1}s^{-6(\mu_{\tau}+1)}, \nonumber \\
&&r_{\mu_{\tau}+1}=-\bar{r}_{\mu_{\tau}+1}\beta \tilde{\Phi}(0), ~~
	\bar{r}_{\mu_{\tau}+1}=2f_0,  \\
&&u_{\mu_{\tau}+1}=\bar{u}_{\mu_{\tau}+1} (\beta \tilde{\Phi}(0))^2, 
	~~\bar{u}_{\mu_{\tau}+1}=\varphi_0-f_0\varphi_0^{-1/2} w_{21}^{(0)}, 
\nonumber \\
&&w_{\mu_{\tau}+1}=\bar{w}_{\mu_{\tau}+1} (\beta \tilde{\Phi}(0))^3, 
	~~\bar{w}_{\mu_{\tau}+1}=\psi_0-f_0\varphi_0 w_{31}^{(0)}. \nonumber
\eee 
Let us perform the change of variables 
\be
\rho_{\vec{k}}=\rho_{\vec{k}}^{'}+\sqrt{N}<\bar{\sigma}>\delta_{\vec{k}}
\ee
in the expression for $Z_{\mu_{\tau}+1}$ in order to extract the free 
energy related to the ordering that has appeared in the system.
Here $<\bar{\sigma}>$ is being determined from the extremum condition
[17,10]
\bee
<\bar{\sigma}>^2&=&10\frac{a_4^{(\mu_{\tau}+1)}}{a_6^{(\mu_{\tau}+1)}}\frac{N_
{\mu_{\tau}+1}}{N}(-1+b_2), \\ 
b_2&=&\sqrt{1+\frac{6a_6^{(\mu_{\tau}+1)}
\mid d_{\mu_{\tau}+1}(0)\mid}{5(a_4^{(\mu_{\tau}+1)})^2}}.\nonumber
\eee 
Simultaneously, we include in the treatment a constant external 
field $h=\mu_{B}{\cal H}$, which sustains the separated average moment, 
and separate from the sums over $k$ the terms with $k=0$. We obtain
\bee
&&Z_{\mu_{\tau}+1}=\exp(-\beta F_{\sigma}+\beta F_{h})\int d\rho_{0}
	\exp \{ \beta \sqrt{N} h \rho_{0}-\frac{1}{2}\bar{d}_{\mu_{\tau}+1}
	(0) \rho_0^2-  \nonumber \\
&&-\frac{b_3^{(\mu_{\tau}+1)}}{3!\sqrt{N_{\mu_{\tau}+1}}}\rho_0^3-
	\frac{b_4^{(\mu_{\tau}+1)}}{4!N_{\mu_{\tau}+1}}\rho_0^4-
	\frac{b_5^{(\mu_{\tau}+1)}}{5!N_{\mu_{\tau}+1}\sqrt{N_{\mu_{\tau}+1}}}
	\rho_0^5-\frac{a_6^{(\mu_{\tau}+1)}}{6!N_{\mu_{\tau}+1}^2}\rho_0^6 
	\}\times\nonumber \\
&&\times \int \exp \{ -\frac{1}{2}\slip_{k\leq 
	B_{\mu_{\tau}+1}}\bar{d}_
	{\mu_{\tau}+1}(k)\rho_{\vec{k}}\rho_{-\vec{k}}\}\exp \{ p_0+p_1\rho_0+
	p_2\rho_0^2+ \nonumber \\
&&+p_3\rho_0^3+p_4\rho_0^4 \} (d\rho)^{N_{\mu_{\tau}+1}-1}.
\eee 
The prime on the sum over $k$ means that $k \neq 0$,
\bee
&&-\beta F_{\sigma}=\frac{10}{3}\mid d_{\mu_{\tau}+1}(0)\mid \frac{a_4^
	{(\mu_{\tau}+1)}}{a_6^{(\mu_{\tau}+1)}}N_{\mu_{\tau}+1}(-1+b_2)-  
	\nonumber \\
&&-\frac{25}{18}\frac{(a_4^{(\mu_{\tau}+1)})^3}{(a_6^{(\mu_{\tau}+1)})^2}
	N_{\mu_{\tau}+1}(-1+b_2)^2, \\
&&\beta F_{h}=\beta \sqrt{N} h \frac{\sqrt{10a_4^{(\mu_{\tau}+1)}} \sqrt{N_
	{\mu_{\tau}+1}}}{\sqrt{a_6^{(\mu_{\tau}+1)}}}(-1+b_2)^{1/2}. \nonumber
\eee
We have for the integrand coefficients of (3.16)
\bee
&&\bar{d}_{\mu_{\tau}+1}(k)= 4\mid d_{\mu_{\tau}+1}(0)\mid 
	-\frac{10}{3}\frac{(a_4^{(\mu_{\tau}+1)})^2}{a_6^{(\mu_{\tau}+1)}}(-1+b_2)+\tilde{q}k^2= 
	\nonumber\\
&&=c_{\nu}^2\mid\tau\mid^{2\nu}\beta\tilde{\Phi}(0)[4\bar{r}_{\mu_{\tau}+1}-\frac{10}{3}\frac{\bar{u}_{\mu_{\tau}+1}^2}{\bar{w}_{\mu_{\tau}+1}}(-1+b_2)]+
	\tilde{q}k^2,  \\
&&b_3^{(\mu_{\tau}+1)}=\frac{\sqrt{10}a_4^{(\mu_{\tau}+1)}\sqrt{a_4^{(\mu_{\tau}
	+1)}}}{\sqrt{a_6^{(\mu_{\tau}+1)}}}(-1+b_2)^{1/2}[1+\frac{5}{3}(-1+b_2)], 
	\nonumber \\
&&b_4^{(\mu_{\tau}+1)}=a_4^{(\mu_{\tau}+1)}[1+5(-1+b_2)], \nonumber\\
&&b_5^{(\mu_{\tau}+1)}=\sqrt{10}\sqrt{a_4^{(\mu_{\tau}+1)}a_6^{(\mu_{\tau}+1)}}
	(-1+b_2)^{1/2}. \nonumber
\eee
The quantities $p_{l}$ are being determined by the equations
\bee
&&p_0=-\frac{b_3^{(\mu_{\tau}+1)}}{3!\sqrt{N_{\mu_{\tau}+1}}}
	\slip_{k_{i}\leq B_{\mu_{\tau}+1}}
	\rho_{\vec{k_1}}\rho_{\vec{k_2}}\rho_{\vec{k_3}}
	\delta_{\vec{k}_1+\vec{k}_2+\vec{k}_3}-\frac{b_4^{(\mu_{\tau}+1)}}
	{4!N_{\mu_{\tau}+1}}\times \nonumber \\
&&\times \slip_{k_{i}\leq B_{\mu_{\tau}+1}}\rho_{\vec{k}_1} \cdots
	\rho_{\vec{k}_4}\delta_{\vec{k}_1+\cdots+\vec{k}_4}-
	\frac{b_5^{(\mu_{\tau}+1)}}{5!N_{\mu_{\tau}+1}\sqrt{N_{\mu_{\tau}+1}}}
	\times \nonumber \\
&&\times\slip_{k_{i}\leq B_{\mu_{\tau}+1}}\rho_{\vec{k}_1}\cdots \rho_
	{\vec{k}_5}\delta_{\vec{k}_1+\cdots+\vec{k}_5}-\frac{a_6^{(\mu_{\tau}+1)}}
	{6!N_{\mu_{\tau}+1}^2}\slip_{k_{i}\leq B_{\mu_{\tau}+1}} 
	\rho_{\vec{k}_1}\cdots\rho_{\vec{k}_6}\delta_{\vec{k}_1+\cdots+\vec{k}_6},
	\nonumber \\ 
&&p_1=-\frac{b_3^{(\mu_{\tau}+1)}}{2\sqrt{N_{\mu_{\tau}+1}}}
	\slip_{k\leq B_{\mu_{\tau}+1}}\rho_{\vec{k}}
	\rho_{-\vec{k}}-\frac{b_4^{(\mu_{\tau}+1)}}{3!N_{\mu_{\tau}+1}}
	\slip_{k_{i}\leq B_{\mu_{\tau}+1}}\rho_{\vec{k}_1}\rho_{\vec{k}_2}\rho_
	{\vec{k}_3} \times \nonumber \\
&&\times\delta_{\vec{k}_1+\vec{k}_2+\vec{k}_3}-\frac{b_5^{(\mu_{\tau}+1)}}
	{4!N_{\mu_{\tau}+1}\sqrt{N_{\mu_{\tau}+1}}}
	\slip_{k_{i}\leq B_{\mu_{\tau}+1}} 
	\rho_{\vec{k}_1}\cdots\rho_{\vec{k}_4}\delta_{\vec{k}_1+\cdots+\vec{k}_4}-
	\nonumber \\
&&- \frac{a_6^{(\mu_{\tau}+1)}}{5!{N_{\mu_{\tau}+1}^2}}
	\slip_{k_{i}\leq B_{\mu_{\tau}+1}} 
	\rho_{\vec{k}_1}\cdots\rho_{\vec{k}_5}\delta_{\vec{k}_1+\cdots+\vec{k}_5},
	\\
&&p_2=-\frac{b_4^{(\mu_{\tau}+1)}}{4N_{\mu_{\tau}+1}}
	\slip_{k\leq B_{\mu_{\tau}+1}}\rho_{\vec{k}}\rho_{-\vec{k}}-
	\frac{b_5^{(\mu_{\tau}+1)}}{12N_{\mu_{\tau}+1}\sqrt{N_{\mu_{\tau}+1}}} 
	\times \nonumber\\
&&\times\slip_{k_{i}\leq 
	B_{\mu_{\tau}+1}}\rho_{\vec{k}_1}\rho_{\vec{k}_2}\rho_{\vec{k}_3} 
	\delta_{\vec{k}_1+\vec{k}_2+\vec{k}_3}-\frac{a_6^{(\mu_{\tau}+1)}}{48{N_{\mu_{\tau}+1}^2}}\slip_{k_{i}\leq 
	B_{\mu_{\tau}+1}} 
	\rho_{\vec{k}_1}\cdots\rho_{\vec{k}_4}\delta_{\vec{k}_1+\cdots+\vec{k}_4},\nonumber 
	\\
&&p_3=-\frac{b_5^{(\mu_{\tau}+1)}}{12N_{\mu_{\tau}+1}\sqrt{N_{\mu_{\tau}+1}}}\slip_{k\leq 
	B_{\mu_{\tau}+1}}\rho_{\vec{k}}\rho_{-\vec{k}}-\frac{a_6^{(\mu_{\tau}+1)}}
	{36N_{\mu_{\tau}+1}^2}\times\nonumber \\
&&\times\slip_{k_{i}\leq 
	B_{\mu_{\tau}+1}}\rho_{\vec{k}_1}\rho_{\vec{k}_2}\rho_{\vec{k}_3} 
	\delta_{\vec{k}_1+\vec{k}_2+\vec{k}_3},\nonumber \\
&&p_4=-\frac{a_6^{(\mu_{\tau}+1)}}{48N_{\mu_{\tau}+1}^2}
	\slip_{k\leq B_{\mu_{\tau}+1}}\rho_{\vec{k}}\rho_{-\vec{k}}.
	\nonumber
\eee
The dependence of the quantity 
$B_6=4\bar{r}_{\mu_{\tau}+1}-\frac{10}{3}\frac{\bar{u}_{\mu_{\tau}+1}^2}
{\bar{w}_{\mu_{\tau}+1}}(-1+b_2)$ on the parameter $s$ is plotted in figure 
3. The dashed line shows the dependence of the analogous quantity $4f_0$ on 
$s$ for the $\rho^4$ model.
\begin{figure}[htb]
\epsfxsize 85mm
\epsfysize 45mm
\centerline{\epsffile{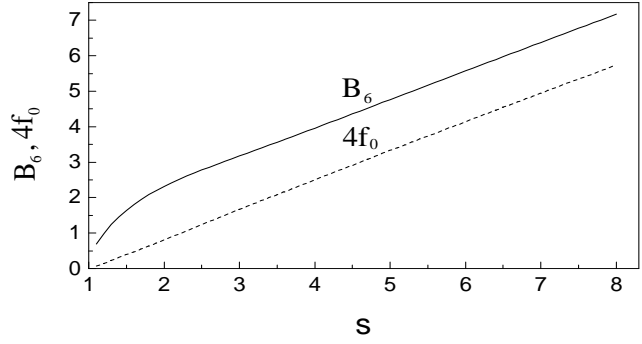}}
\caption {Dependence of the quantities 
$B_6=4\bar{r}_{\mu_{\tau}+1}-\frac{10}{3}\frac{\bar{u}_{\mu_{\tau}+1}^2}
{\bar{w}_{\mu_{\tau}+1}}(-1+b_2)$ ($\rho^6$ model) and $4f_0$ ($\rho^4$ 
model) on $s$.}
\end{figure}
One can see from figure 3 the quantity $B_6$ ocurring in 
$\bar{d}_{\mu_{\tau}+1}(k)$ (3.18) (therefore, and 
$\bar{d}_{\mu_{\tau}+1}(k)$) is positive for all $s$.

Expanding $\exp \{p_0+p_1\rho_0+p_2\rho_0^2+p_3\rho_0^3+p_4\rho_0^4 \}$ 
in series and restricting ourselves to the terms of second order, we 
integrate (3.16) over $\rho_{\vec{k}}$ with $k\neq 0$ using the 
Gaussian basis distribution. Gathering up the series over the averages 
with respect to the Gaussian distribution in the exponential, we obtain [15]
\bee
&&Z_{\mu_{\tau}+1}=\exp(-\beta F_{\sigma}+\beta F_{h}-\beta F_{m})\plip_
	{k_{i}\leq 
	B_{\mu_{\tau}+1}}(\frac{\pi}{\bar{d}_{\mu_{\tau}+1}(k)})^{1/2}\times\nonumber 
	\\
&&\times\int\exp(\tilde{A}_1\rho_0+\tilde{A}_2\rho_0^2+\tilde{A}_3\rho_0^3+
        \tilde{A}_4\rho_0^4+\tilde{A}_5\rho_0^5+\tilde{A}_6\rho_0^6+\nonumber 
        \\
&&+\tilde{A}_7\rho_0^7+\tilde{A}_8\rho_0^8)d\rho_0,
\eee
where
\bee
-\beta 
F_{m}&=&\frac{1}{4}N_{\mu_{\tau}+1}[-\frac{1}{2}(b_4^{(\mu_{\tau}+1)}+\frac{a_6^{(\mu_{\tau}+1)}}{6}{\cal 
	I}_1){\cal I}_1^2+\frac{b_3^{(\mu_{\tau}+1)}}{3}{\cal I}_3 
	\times \nonumber \\
&&\times (b_3^{(\mu_{\tau}+1)}+b_5^{(\mu_{\tau}+1)}{\cal I}_1)+
	\frac{(b_4^{(\mu_{\tau}+1)})^2}{4}(\frac{{\cal I}_4}{3}+{\cal 
	I}_1^2{\cal I}_2)+\frac{(b_5^{(\mu_{\tau}+1)})^2}
	{12}\times \nonumber \\
&& \times
	(\frac{{\cal I}_5}{5}+{\cal I}_1^2{\cal 
	I}_3)+\frac{(a_6^{(\mu_{\tau}+1)})^2}
	{8}(\frac{{\cal I}_6}{45}+\frac{{\cal I}_1^2}{2}(\frac{{\cal 
	I}_1^2{\cal I}_2}{4}+\frac{{\cal I}_4}{3}))+\nonumber\\
&&+\frac{b_4^{(\mu_{\tau}+1)}a_6^{(\mu_{\tau}+1)}}{4}
	{{\cal I}_1}(\frac{{\cal I}_4}{3}+\frac{{\cal I}_1^2{\cal 
	I}_2}{2})], \nonumber \\
\tilde{A}_1&=&\beta\sqrt{N}h+\frac{1}{2}N_{\mu_{\tau}+1}^{1/2}[-{\cal I}_1
	(b_3^{(\mu_{\tau}+1)}+\frac{1}{4}b_5^{(\mu_{\tau}+1)}{\cal I}_1)+
	\nonumber \\
&&+b_3^{(\mu_{\tau}+1)}b_4^{(\mu_{\tau}+1)}(\frac{{\cal 
	I}_3}{3}+\frac{{\cal I}_1{\cal 
	I}_2}{2})+\frac{b_3^{(\mu_{\tau}+1)}a_6^{(\mu_{\tau}+1)}}{2}{\cal I}_1
	(\frac{{\cal I}_3}{3}+\frac{{\cal I}_1{\cal I}_2}{4})+ \nonumber \\
&&+\frac{b_4^{(\mu_{\tau}+1)}b_5^{(\mu_{\tau}+1)}}{2}
	(\frac{{\cal I}_4}{6}+{\cal I}_1(\frac{{\cal I}_3}{3}+\frac{{\cal 
	I}_1{\cal 
	I}_2}{2}))+ \frac{b_5^{(\mu_{\tau}+1)}a_6^{(\mu_{\tau}+1)}}{4}
	(\frac{{\cal I}_5}{15}+ \nonumber \\
&&+{\cal I}_1(\frac{{\cal I}_1{\cal I}_3}{3}+
	\frac{{\cal I}_4}{6}+\frac{{\cal I}_1^2{\cal I}_2}{4}))], \nonumber \\
\tilde{A}_2&=&-2\mid d_{\mu_{\tau}+1}(0)\mid + 
	\frac{5}{3}\frac{(a_4^{(\mu_{\tau}+1)})^2}{a_6^{(\mu_{\tau}+1)}}(-1+b_2)+
	\frac{1}{4}[-{\cal I}_1(b_4^{(\mu_{\tau}+1)}+ \nonumber \\
&&+\frac{a_6^{(\mu_{\tau}+1)}}{4}{\cal I}_1)+ 
	(b_3^{(\mu_{\tau}+1)})^2{\cal I}_2+(b_4^{(\mu_{\tau}+1)})^2(\frac{{\cal 
	I}_3}{3}+\frac{{\cal I}_1{\cal 
	I}_2}{2})+\nonumber \\
&&+\frac{(b_5^{(\mu_{\tau}+1)})^2}{2}
	(\frac{{\cal I}_4}{6}+{\cal I}_1(\frac{{\cal I}_3}{3}+\frac{{\cal 
	I}_1{\cal I}_2}{2}))+\frac{(a_6^{(\mu_{\tau}+1)})^2}{4}
	(\frac{{\cal I}_5}{15}+\nonumber \\
&&+{\cal I}_1(\frac{{\cal I}_1{\cal 
	I}_3}{3}+\frac{{\cal I}_4}{6}+\frac{{\cal I}_1^2{\cal 
	I}_2}{4}))+b_3^{(\mu_{\tau}+1)}b_5^{(\mu_{\tau}+1)}(\frac{{\cal I}_3}{3}+
	{\cal I}_1{\cal 
	I}_2)+\nonumber \\
&&+b_4^{(\mu_{\tau}+1)}a_6^{(\mu_{\tau}+1)}
	(\frac{{\cal I}_4}{12}+{\cal I}_1(\frac{{\cal I}_3}{3}+\frac{3}{8}
	{\cal I}_1{\cal I}_2))],  \\
\tilde{A}_3&=&\frac{1}{2N_{\mu_{\tau}+1}^{1/2}}[-\frac{1}{3}(b_3^{(\mu_{\tau}+1)}+
	\frac{b_5^{(\mu_{\tau}+1)}}{2}{\cal 
	I}_1)+\frac{b_3^{(\mu_{\tau}+1)}a_6^{(\mu_{\tau}+1)}}{2}(\frac{{\cal 
	I}_3}{9}+ \nonumber \\
&&+\frac{{\cal I}_1{\cal 
	I}_2}{2})+\frac{b_4^{(\mu_{\tau}+1)}b_5^{(\mu_{\tau}+1)}}{2}(\frac{{\cal 
	I}_3}{3}+\frac{2}{3}{\cal I}_1{\cal I}_2)+
	\frac{b_5^{(\mu_{\tau}+1)}a_6^{(\mu_{\tau}+1)}}{12}(\frac{{\cal 
	I}_4}{2}+ \nonumber \\
&&+{\cal I}_1(\frac{4{\cal I}_3}{3}+\frac{7}{4}{\cal I}_1{\cal I}_2))
	+\frac{b_3^{(\mu_{\tau}+1)}b_4^{(\mu_{\tau}+1)}}{2}{\cal I}_2], \nonumber 
	\\
\tilde{A}_4&=&\frac{1}{4N_{\mu_{\tau}+1}}[-\frac{1}{6}(b_4^{(\mu_{\tau}+1)}+
	\frac{a_6^{(\mu_{\tau}+1)}}{2}{\cal I}_1)+\frac{b_4^{(\mu_{\tau}+1)}}{4}
	{\cal I}_2(b_4^{(\mu_{\tau}+1)}+ \nonumber \\
&&+\frac{7}{6}a_6^{(\mu_{\tau}+1)}{\cal 
	I}_1)+\frac{(b_5^{(\mu_{\tau}+1)})^2}{6}(\frac{{\cal I}_3}{2}+{\cal I}_1 
	{\cal I}_2)+\frac{b_3^{(\mu_{\tau}+1)}b_5^{(\mu_{\tau}+1)}}{3}{\cal 
	I}_2+ \nonumber \\
&&+\frac{a_6^{(\mu_{\tau}+1)}b_4^{(\mu_{\tau}+1)}}{9}{\cal I}_3+
	\frac{(a_6^{(\mu_{\tau}+1)})^2}{6}(\frac{{\cal I}_4}{8}+{\cal I}_1(
	\frac{{\cal I}_3}{3}+\frac{7}{16}{\cal I}_1{\cal I}_2))], \nonumber \\
\tilde{A}_5&=&\frac{1}{24N_{\mu_{\tau}+1}^{3/2}}[b_5^{(\mu_{\tau}+1)}
	(-\frac{1}{5}+b_4^{(\mu_{\tau}+1)}{\cal 
	I}_2)+\frac{a_6^{(\mu_{\tau}+1)}{\cal I}_2}{2}(b_3^{(\mu_{\tau}+1)}+
	\nonumber \\
&&+\frac{b_5^{(\mu_{\tau}+1)}}{2}{\cal 
	I}_1)+b_5^{(\mu_{\tau}+1)}a_6^{(\mu_{\tau}+1)}(\frac{{\cal 
	I}_3}{3}+\frac{{\cal I}_1{\cal I}_2}{3})], \nonumber \\
\tilde{A}_6&=&\frac{1}{48N_{\mu_{\tau}+1}^2}[\frac{a_6^{(\mu_{\tau}+1)}}{3}	
	(-\frac{1}{5}+
	\frac{a_6^{(\mu_{\tau}+1)}}{3}{\cal I}_3)+
	\frac{(b_5^{(\mu_{\tau}+1)})^2}{3}{\cal I}_2+ \nonumber \\
&&+\frac{a_6^{(\mu_{\tau}+1)}}{2}
	{\cal I}_2(b_4^{(\mu_{\tau}+1)}+\frac{a_6^{(\mu_{\tau}+1)}}{2}{\cal 
	I}_1)], \nonumber \\
\tilde{A}_7&=&\frac{b_5^{(\mu_{\tau}+1)}a_6^{(\mu_{\tau}+1)}}{288N_{\mu_
	{\tau}+1}^{5/2}}{\cal I}_2, \nonumber \\
\tilde{A}_8&=&\frac{(a_6^{(\mu_{\tau}+1)})^2}{2304N_{\mu_{\tau}+1}^{3}}{\cal 
	I}_2. \nonumber 
\eee

Expressions for ${\cal I}_m$ occurring in (3.21) can be found  via the 
transition to the spherical Brillouin zone and integration over 
$k \in (0,B_{\mu_{\tau}+1}]$ [17,18].

We have
\bee
{\cal I}_1&=&\frac{1}{N_{\mu_{\tau}+1}}\slip_{k \leq 
	B_{\mu_{\tau}+1}}\frac{1}{\bar{d}_{\mu_{\tau}+1}(k)}={\cal L}[4\mid 
	d_{\mu_{\tau}+1}(0)\mid -\frac{10}{3}\times \nonumber \\
&&\times \frac{(a_4^{(\mu_{\tau}+1)})^2}{a_6^{(\mu_{\tau}+1)}}(-1+b_2)]^{-1}
\nonumber 
\eee 
or
\be
{\cal I}_1=\frac{s^{2(\mu_{\tau}+1)}}{\beta\tilde{\Phi}(0)}\alpha_1. 
\nonumber 
\ee
Here,
\bee
&&{\cal L}=3\frac{x_r-\arctan{x_r}}{x_r^3}, \quad
x_r=\frac{1}{\sqrt{4\bar{r}_{\mu_{\tau}+1}-\frac{10}{3}\frac{\bar{u}_{\mu_{\tau}+1}^2}
{\bar{w}_{\mu_{\tau}+1}}(-1+b_2)}}, \\
&& b_2=\sqrt{1+\frac{6}{5}\frac{\bar{w}_{\mu_{\tau}+1}\bar{r}_{\mu_{\tau}+1}
}{\bar{u}_{\mu_{\tau}+1}^2}},\quad 
\alpha_1=\frac{{\cal L}}{4\bar{r}_{\mu_{\tau}+1}-\frac{10}{3}\frac{\bar{u}_{\mu_{\tau}+1}^2}
{\bar{w}_{\mu_{\tau}+1}}(-1+b_2)}. \nonumber
\eee
The quantity ${\cal I}_2$ can be represented in the form:
\be
{\cal I}_2=\sum_r g^2(r)=\frac{s^{4(\mu_{\tau}+1)}}{(\beta 
\tilde{\Phi}(0))^2}\alpha_2,
\ee
where
\bee
g(r)&=&\frac{1}{N_{\mu_{\tau}+1}}\slip_{k \leq 
	B_{\mu_{\tau}+1}}\frac{e^{i 
	\vec{k}\vec{r}}}{\bar{d}_{\mu_{\tau}+1}(k)}=\frac{6s^{2(\mu_{\tau}+1)}}{\beta 
	\tilde{\Phi}(0)(B_{\mu_{\tau}+1}r)^3}\{ 
	8\bar{r}_{\mu_{\tau}+1} -\nonumber \\
&&-\frac{20}{3}\frac{\bar{u}_{\mu_{\tau}+1}^2}
	{\bar{w}_{\mu_{\tau}+1}}(-1+b_2)+1 \}^{-1}[\sin(B_{\mu_{\tau}+1}r)-
	B_{\mu_{\tau}+1}r\cos(B_{\mu_{\tau}+1}r)], \nonumber \\
\alpha_2&=&\alpha_1^2+6e_1^2(1+e_2^2), \\
e_1&=&\frac{6}{\pi^2[8\bar{r}_{\mu_{\tau}+1}-\frac{20}{3}\frac{\bar{u}_{\mu_{\tau}+1}^2}
	{\bar{w}_{\mu_{\tau}+1}}(-1+b_2)+1]}, \nonumber \\
e_2&=&\frac{1}{2\pi}[\sin(\pi\sqrt{2})-\pi\sqrt{2}\cos(\pi\sqrt{2})]\approx 
	0.034861. \nonumber 
\eee
The other ${\cal I}_l (l=3,4,5,6)$ are determined by analogous relations
\be
{\cal I}_l=\sum_r g^l(r)=\frac{s^{2l(\mu_{\tau}+1)}}{(\beta 
\tilde{\Phi}(0))^l}\alpha_l
\ee
with
\be
\alpha_l=\alpha_1^l+6e_1^l(1+\frac{e_2^l}{2^{l/2-1}}).
\ee
The values of $\alpha_m (m=1,2,3,4,5,6)$ are given in table 6.
\begin{table}[htb]
\caption{Numerical values of $\alpha_m$.}
\begin{tabular}{lllllll}
&  &  &  &  &  & \\
\hline
&  &  &  &  &  & \\
~~$s$~~&$\alpha_1$~&~$\alpha_2
$~~&~$\alpha_3$~~&$\alpha_4$~&~~$\alpha_5$~~&~~
$\alpha_6$\\
&  &  &  &  &  & \\
\hline
&  &  &  &  &  & \\
2~~~& 0.3456~~~& 0.1893~~~& 0.0488 
~~~& 0.0151~~~& 0.0050~~~& 0.0017 \\
2.5    & 0.2981 & 0.1406 & 0.0313 & 0.0083 & 0.0024 & 0.0007 \\
2.7349 & 0.2821 & 0.1258 &  0.0265 &  0.0067 & 0.0018 & 0.0005 \\ 
3      & 0.2664 & 0.1122 &  0.0223 &  0.0053 & 0.0014 & 0.0004 \\
3.5    & 0.2412 & 0.0918 &  0.0165 &  0.0036 & 0.0008 & 0.0002 \\
3.5862 & 0.2373 & 0.0888 &  0.0158 &  0.0033 & 0.0008 & 0.0002 \\
4      & 0.2201 & 0.0764 &  0.0126 &  0.0025 & 0.0005 & 0.0001 \\
&  &  &  &  &  & \\
\hline
\end{tabular}
\end{table}

Let us perform the change of the variable $\rho_0$ in (3.20)
\be
\rho_0=\rho_0^{'}-\sqrt{N}<\bar{\sigma}>,
\ee
which cancels terms proportional to odd powers of $\rho_0$ in the exponent 
of the integrand. We obtain
\bee
Z_{\mu_{\tau}+1}&=&\exp (-\beta F_{\mu_{\tau}+1}^{'})\int \exp[\beta 
\sqrt{N}\rho_0 h 
+\tilde{B}\rho_0^2-\frac{G}{N}\rho_0^4-
\nonumber \\
&-&\frac{D}{N^2}\rho_0^6]d\rho_0,
\eee
where
\bee
-\beta F_{\mu_{\tau}+1}^{'}&=&N_{\mu_{\tau}+1}\{ 
	\frac{5}{2}\bar{r}_{\mu_{\tau}+1}(\alpha_1+\frac{5}{2}\bar{r}_{\mu_{\tau}+1}\alpha_2)-
	\nonumber \\
&&-\frac{\alpha_1^2}{8}(\bar{u}_{\mu_{\tau}+1}+\frac{\bar{w}_{\mu_{\tau}+1}}{6}
	\alpha_1)+\frac{\bar{u}_{\mu_{\tau}+1}^2}{8}[\frac{\alpha_4}{6}+\alpha_1^2\alpha_2(\frac{1}{2}+
\nonumber \\
	&&+\frac{5}{3}(-1+
	b_2))]-\frac{5}{4}\bar{r}_{\mu_{\tau}+1}\alpha_1\alpha_2[\bar{u}_{\mu_{\tau}+1}
	+\frac{\bar{w}_{\mu_{\tau}+1}}{4}\alpha_1]+\nonumber 
	\\
&&+\frac{\bar{w}_{\mu_{\tau}+1}^2}{32}
	[\frac{\alpha_6}{45}+\frac{\alpha_1^2}{2}(\frac{\alpha_4}{3}+\frac{\alpha_1^
	2\alpha_2}{4})]+\nonumber \\
&&+\frac{\bar{u}_{\mu_{\tau}+1}\bar{w}_{\mu_{\tau}+1}\alpha_1}{16}[
	\frac{\alpha_4}{3}+\frac{\alpha_1^2\alpha_2}{2}]-\nonumber \\
&&-\frac{5}{3}\frac{\bar{u}_{\mu_{\tau}+1}^2}{\bar{w}_{\mu_{\tau}+1}}[
	(-1+b_2)(\alpha_1-\bar{u}_{\mu_{\tau}+1}\alpha_2(\frac{\alpha_1}{2}-\frac{10}{3}
	\frac{\bar{u}_{\mu_{\tau}+1}}{\bar{w}_{\mu_{\tau}+1}}))- \nonumber \\
&&-(7-5b_2)\bar{r}_{\mu_{\tau}+1}\alpha_2]-\frac{1}{2}\ln 
	(\frac{\beta\tilde{\Phi}(0)}{s^{2(\mu_{\tau}+1)}})-\frac{1}{2}\ln 
	[(1+\nonumber \\
&&+ 4\bar{r}_{\mu_{\tau}+1}- 
	\frac{10}{3}\frac{\bar{u}_{\mu_{\tau}+1}^2}{\bar{w}_{\mu_{\tau}+1}}(-1+b_2))\pi^{-1}]+\frac{1}{3}-\frac{{\cal 
	L}}{3}\},\nonumber \\ 
\tilde{B}&=&\frac{1}{2}c_{\nu}^2\mid\tau\mid^{2\nu}\beta\tilde{\Phi}
	(0)\bar{r}_{\mu_{\tau}+1}\bar{B}, \\
G&=&c_{\nu}\mid\tau\mid^{\nu}(\beta\tilde{\Phi}(0))^2s_0^3\bar{G}, 
	\nonumber \\D&=&(\beta\tilde{\Phi}(0))^3s_0^6\bar{D}, \nonumber \\
\bar{B}&=&1-\frac{\alpha_1}{2\bar{r}_{\mu_{\tau}+1}}(\bar{u}_{\mu_{\tau}+1}+\frac{\bar{w}_{\mu_{\tau}+1}}{4}
	\alpha_1)+\frac{\bar{u}_{\mu_{\tau}+1}^2}{\bar{r}_{\mu_{\tau}+1}}[\frac{5}{3}(-1+b_2)
	\alpha_2 \times \nonumber \\
&&\times(\frac{\alpha_1}{2}+\frac{\bar{u}_{\mu_{\tau}+1}}{\bar{w}_{\mu_{\tau}+1}}
	)+\frac{1}{2}(\frac{\alpha_1\alpha_2}{2}+\frac{\alpha_3}{3})]-\frac{5}{2}\alpha_2(
	\bar{u}_{\mu_{\tau}+1}+\nonumber \\
&&+\frac{\bar{w}_{\mu_{\tau}+1}\alpha_1}{2})+\frac{\bar{w}_{\mu_{\tau}+1}^2}{8\bar{r}_{\mu_{\tau}+1}}[\frac{\alpha_5}{15}
	+\alpha_1(\frac{\alpha_1\alpha_3}{3}+\frac{1}{2}(\frac{\alpha_4}{3}+
	\frac{\alpha_1^2\alpha_2}{2}))]+
	\nonumber \\
&&+\frac{\bar{u}_{\mu_{\tau}+1}\bar{w}_{\mu_{\tau}+1}}{2\bar{r}_{\mu_{\tau}+1}}
	[\frac{\alpha_4}{12}+\alpha_1(\frac{3\alpha_1\alpha_2}{8}+\frac{\alpha_3}{3})],
	\nonumber \\
\bar{G}&=&\frac{1}{24}[\bar{u}_{\mu_{\tau}+1}+\frac{\bar{w}_{\mu_{\tau}+1}}{2}
	(\alpha_1+5\bar{r}_{\mu_{\tau}+1}\alpha_2)]-\frac{\bar{u}_{\mu_{\tau}+1}^2
	\alpha_2}{8}[\frac{1}{2}+\frac{5}{9}(-1+ \nonumber \\
&&+b_2)]-\frac{\bar{u}_{\mu_{\tau}+1}\bar{w}_{\mu_{\tau}+1}}{12}(\frac{7}{8}\alpha_1
	\alpha_2+\frac{\alpha_3}{3})-\frac{\bar{w}_{\mu_{\tau}+1}^2}{24}[
	\frac{\alpha_4}{8}+\nonumber \\
&&+\alpha_1(\frac{7}{16}\alpha_1\alpha_2+\frac{\alpha_3}{3})], \nonumber \\
\bar{D}&=&\frac{\bar{w}_{\mu_{\tau}+1}}{48}(\frac{1}{15}-\frac{\bar{u}_
	{\mu_{\tau}+1}}{2}\alpha_2)-\frac{\bar{w}_{\mu_{\tau}+1}^2}{48}(\frac{\alpha_3}{9}+
	\frac{\alpha_1\alpha_2}{4}).\nonumber
\eee
The quantities $\bar{B}, \bar{G}, \bar{D}$ are given in table 7.
\begin{table}[htb]
\caption{Values of $\bar{B}, \bar{G}, \bar{D}, 
<\bar{\sigma}>^{(0)}, \Gamma^+,\Gamma^-$ for different $s$.}
\begin{tabular}{lllllll}
&  &  &  &  &  & \\
\hline
&  &  &  &  &  & \\
~~$s$~~&$\bar B$~&~$\bar G
$~~&~$\bar D$~~&$<\bar{\sigma}>^{(0)}$~&~~$\Gamma^+$~~&~~
$\Gamma^-$\\
&  &  &  &  &  & \\
\hline
&  &  &  &  &  & \\
2~~~& 0.8296~~~& 0.0153~~~& 0.0002 
~~~&  3.0942~~~& 2.6052~~~& 0.2970 \\
2.5    & 0.8289 & 0.0206 & 0.0003  & 3.0031 & 3.8147 & 0.3121 \\
2.7349 & 0.8277 & 0.0231 &  0.0003 &  2.9669 & 2.9709 & 0.3211 \\ 
3      & 0.8267 & 0.0260 &  0.0003 &  2.9309 & 3.3248 & 0.3318 \\
3.5    & 0.8266 & 0.0316 &  0.0003 &  2.8730 & 4.0679 & 0.3528 \\
3.5862 & 0.8268 & 0.0326 &  0.0003 &  2.8640 & 4.2071 & 0.3565 \\
4      & 0.8282 & 0.0376 &  0.0004 &  2.8238 & 4.9254 & 0.3748 \\
&  &  &  &  &  & \\
\hline
\end{tabular}
\end{table}

Using the method of the steepest descent for $Z_{\mu_{\tau}+1}$ (3.29), 
we find
\be
Z_{\mu_{\tau}+1}=\sqrt{\frac{2\pi}{E_0^{''}(\bar{\rho})}}\exp [-\beta 
	F_{\mu_{\tau}+1}^{'}-N E_0(\bar{\rho})].
\ee
Here $\bar{\rho}$ is the extremum point of the expression
\be
E_0(\rho)=D\rho^6+G\rho^4-\tilde{B}\rho^2-\beta h \rho
\ee
arising in the exponential of the integrand of (3.29) at the substitution
\be
\rho_0=\sqrt{N}\rho.
\ee
For $E_0(\bar{\rho})$ at $h=0$ we obtain
\bee
E_0(\bar{\rho})=-s^{-3(\mu_{\tau}+1)}s_0^{-3}\bar{E}_0, \nonumber
\eee
\bee
\bar{E}_0&=&\frac{2\bar{G}^3}{27\bar{D}^2}(-1+\sqrt{1+\frac{3}{2}\frac{\bar{r}_
{\mu_{\tau}+1}\bar{B}\bar{D}}{\bar{G}^2}})+\frac{\bar{r}_{\mu_{\tau}+1}\bar{B}
\bar{G}}{6\bar{D}}\times \\
&&\times (-1+\frac{2}{3}\sqrt{1+\frac{3}{2}\frac{\bar{r}_
{\mu_{\tau}+1}\bar{B}\bar{D}}{\bar{G}^2}}). \nonumber
\eee

Having (3.12) and (3.31), we can calculate the contribution to the system 
free energy at $T<T_c$ from the long-wave phases of the spin moment 
density fluctuations (3.9):
\bee
F_{IGR}&=&-kTN^{'}(\gamma_3^{(\mu_{\tau})}+\gamma_3^{< \sigma >})\mid 
	\tau \mid^{3\nu}, \nonumber \\
\gamma_3^{(\mu_{\tau})}&=&\gamma_g+\gamma_{\rho}=c_{\nu}^3\bar{\gamma_3}^{(\mu_{\tau})},
	\quad \bar{\gamma}_3^{(\mu_{\tau})}=\bar{\gamma}_g+\bar{\gamma}_{\rho}, 
	\nonumber \\
\gamma_{\rho}&=&c_{\nu}^3\bar{\gamma_{\rho}}, \quad 
	\bar{\gamma}_{\rho}=\frac{5}{2}\bar{r}_{\mu_{\tau}+1}(\alpha_1+\frac{5}{2}\bar{r}
	_{\mu_{\tau}+1}\alpha_2)-\frac{\alpha_1^2}{8} \times \nonumber \\
&& \times (\bar{u}_{\mu_{\tau}+1}+\frac{\bar{w}_{\mu_{\tau}+1}}{6}\alpha_1)+
	\frac{\bar{u}_{\mu_{\tau}+1}^2}{8}[\frac{\alpha_4}{6}+\alpha_1^2\alpha_2(\frac{1}{2}+
	\frac{5}{3}(-1+b_2))]- \nonumber \\
&&-\frac{5}{4}\bar{r}_{\mu_{\tau}+1}\alpha_1\alpha_2[\bar{u}_{\mu_{\tau}+1}+
	\frac{\bar{w}_{\mu_{\tau}+1}}{4}\alpha_1]+\frac{\bar{w}_{\mu_{\tau}+1}^2}{32}[
	\frac{\alpha_6}{45}+\frac{\alpha_1^2}{2} \times \\
&&\times 
	(\frac{\alpha_4}{3}+\frac{\alpha_1^2\alpha_2}{4})]+\frac{\bar{u}_
	{\mu_{\tau}+1}\bar{w}_{\mu_{\tau}+1}\alpha_1}{16}[\frac{\alpha_4}{3}+\frac{
	\alpha_1^2\alpha_2}{2}]-\frac{5}{3}\frac{\bar{u}_{\mu_{\tau}+1}^2}{\bar{w}_{
	\mu_{\tau}+1}}\times \nonumber \\
&&\times 
	[(-1+b_2)(\alpha_1-\bar{u}_{\mu_{\tau}+1}\alpha_2(\frac{\alpha_1}{2}-\frac{10}
	{3}\frac{\bar{u}_{\mu_{\tau}+1}}{\bar{w}_{\mu_{\tau}+1}}))-(7-5b_2) 
	\times \nonumber \\
&&\times \bar{r}_{\mu_{\tau}+1}\alpha_2]-\frac{1}{2}\ln [\frac{1+
	4\bar{r}_{\mu_{\tau}+1}-\frac{10}{3}\frac{\bar{u}_{\mu_{\tau}+1}^2}
	{\bar{w}_{\mu_{\tau}+1}}(-1+b_2)}{\pi}]+\frac{1}{3}-\frac{{\cal L}}{3},
	\nonumber \\
\gamma_3^{<\sigma >}&=&c_{\nu}^3\bar{\gamma_3}^{<\sigma >},
	\quad \bar{\gamma_3}^{<\sigma >}=\bar{E}_0. \nonumber 
\eee 
The quantity $\gamma_3^{\mu_{\tau}}$ determines the free energy 
after the exit from the CR region, and $\gamma_3^{<\sigma>}$ 
determines the free energy of the ordering. The values of $\bar{\gamma}_g, 
\bar{\gamma}_{\rho}, \bar{\gamma}_3^{(\mu_{\tau})}, 
\bar{\gamma}_3^{<\sigma >}$ are given in table 8.
\begin{table}[htb]
\caption{Values of $\bar{\gamma}_g, \bar{\gamma}_{\rho},
\bar{\gamma}_3^{(\mu_{\tau})}, \bar{\gamma}_3^{<\sigma>}$.}
\begin{center}
\begin{tabular}{lrlll}
&  &  &  &   \\
\hline
&  &  &  &   \\
~~$s$~~& $\bar{\gamma}_g~~~$ &~~~$\bar{\gamma}_{\rho}
$~~& ~$\bar{\gamma}_3^{(\mu_{\tau})}$~~ & $\bar{\gamma}_3^{<\sigma>}$ \\
&  &  &  &   \\
\hline
&  &  &  &   \\
2~~~~~& -0.3024 & 1.0386~~~~~ & 0.7362 
~~~~~&  1.7618 \\
2.5    & -0.0869 & 1.0269 & 0.9399 & 2.0456  \\
2.7349 & 0.0039  & 1.0227 &  1.0265 &  2.1572  \\ 
3      & 0.0986  & 1.0179 &  1.1164 &  2.2808  \\
3.5    & 0.2579  & 1.0076 &  1.2655 &  2.5154  \\
3.5862 & 0.2832  & 1.0056 &  1.2888 &  2.5563 \\
4      & 0.3967  & 0.9955 &  1.3922 &  2.7538  \\
&  &  &  &   \\
\hline
\end{tabular}
\end{center}
\end{table}

The entropy, internal energy and specific heat of the system
correspon\-ding to the IGR region read
\bee
&&S_{IGR}=S_{\mu_{\tau}}+S_{< \sigma >},\quad 
	U_{IGR}=U_{\mu_{\tau}}+U_{< \sigma >},\quad
	C_{IGR}=C_{\mu_{\tau}}+C_{< \sigma >},\nonumber \\
&&S_{\eta}=-kN^{'}\mid \tau \mid ^{1-\alpha}u_3^{(\eta)}, \quad
	U_{\eta}=-kTN^{'}\mid \tau \mid ^{1-\alpha}u_3^{(\eta)}, \\
&&C_{\eta}=kN^{'}c_3^{(\eta)}\mid \tau \mid ^{-\alpha}, \quad
	u_3^{(\eta)}=3\nu\gamma_3^{(\eta)}, \quad
	c_3^{(\eta)}=3\nu(3\nu-1)\gamma_3^{(\eta)}. \nonumber
\eee  
The index $\eta$ can take two values: $\mu_{\tau}$ ¨ $< \sigma >$.

\sect {The order parameter of the system}

The mean spin moment is the order parameter of the model investigated. 
It is related to the presence of the nonzero value of the 
CV $\rho_0$ at which the integrand of (3.29) has an extremum. 
Performing the substitution (3.33) in that integrand, we obtain
\be
Z_{\mu_{\tau}+1}=e^{-\beta F_{\mu_{\tau}+1}^{'}}\sqrt{N}\int 
e^{-N E_0(\rho)}d\rho, 
\ee
where $E_0(\rho)$ is given in (3.32). Owing to the factor $N$ in 
the exponent in (4.1), the integrand has a sharp maximum at the point $\bar 
\rho $ corresponding to the equilibrium value of the order parameter. The 
value of $\bar \rho $ can be found from the extremum condition 
$\frac{\partial  E_0(\rho)}{\partial \rho}=0$ or
\be
6D\bar{\rho}^5+4G\bar{\rho}^3-2\tilde{B}\bar{\rho}-\beta h=0.
\ee
In the case $h=0$ we obtain a biquadratic equation, which is reduced by 
means of substitution
\be
\bar{\rho}^2=y
\ee
into the equation
\be
6Dy^2+4Gy-2\tilde{B}=0.
\ee
Extracting the temperature dependence, we obtain the equation for the mean 
spin moment $< \sigma >=\bar{\rho}=\sqrt{y}$:
\bee
< \sigma >&=&\mid \tau \mid ^{\beta}< \sigma >^{(0)}, \quad \beta=\nu/2, 
	\nonumber \\
<\sigma >^{(0)}&=&c_{\nu}^{1/2}(\beta\tilde{\Phi}(0))^
	{-1/2}s_0^{-3/2}<\bar{\sigma} >^{(0)}, \\
<\bar{\sigma}>^{(0)}&=&[\frac{\bar{G}}{3\bar{D}}(-1+\sqrt{1+
	\frac{3}{2}\frac{\bar{r}_{\mu_{\tau}+1}\bar{B}\bar{D}}{\bar{G}^2}})]
	^{1/2}. \nonumber
\eee
The value of $< \bar{\sigma} >^{(0)}$ is given in table 7.

The susceptibility per particle $\chi$ can be found from equation 
(4.2) by differentiating it with respect to ${\cal H}$ and using the 
relation	$\chi=\mu_B\frac{\partial <\sigma >}{\partial {\cal H}}$:
\be
\chi=\frac{\beta\mu_B^2}{30 D\bar{\rho}^4 +12G\bar{\rho}^2
-2\tilde{B}}.
\ee 
Separating the temperature dependence in the coefficients 
$D$, $G$, $\tilde{B}$ (see (3.30)), we obtain a final expression for the 
susceptibility.

\sect{Thermodynamics of the system in the vicinity of the phase transition 
point}

Having calculated the contributions to the system free energy from the 
short-wave and long-wave modes of the spin moment density 
oscillations both above and below $T_c$, we can compute the 
total free energy\bee
F &=& 
\left \{
\begin{array}{l}
F_0+F_{CR}+F_{LGR}, \quad T>T_c, \\
F_0+F_{CR}+F_{IGR}, \quad T<T_c, \\
\end{array}
\right . \nonumber 
\eee 
the entropy, internal energy and specific heat.

We obtain the total free energy of the system at $h=0$ taking (2.17),
(2.23) and (3.6), (3.35) into account: 
\bee
F &=& 
\left \{
\begin{array}{l}
-kTN^{'}[\gamma_0+\gamma_1\tau+\gamma_2\tau^2+\gamma_3^{+}\tau^{3\nu}], 
\quad T>T_c, \\
-kTN^{'}[\gamma_0+\gamma_1\mid\tau\mid+\gamma_2\mid\tau\mid^2+\gamma_3^{-}\mid
\tau\mid^{3\nu}], 
\quad T<T_c, 
\end{array}
\right .
\eee 
where (see table 9)
\bee
&&\gamma_0=\gamma_0^{(CR)}+s_0^3\ln2, \nonumber \\
&&\gamma_3^{+}=-\gamma_3^{(CR)+}+f_{LGR}, \\
&&\gamma_3^{-}=-\gamma_3^{(CR)-}+\gamma_{IGR}, \quad 
\gamma_{IGR}=\gamma_3^{(\mu_{\tau})}+\gamma_3^{< \sigma >}. \nonumber 
\eee
\begin{table}[htb]
\caption{Coefficients $\gamma_0$, 
$\gamma_3^{\pm}$ and $\bar{\gamma}_3^{\pm}$.}
\begin{center}
\begin{tabular}{llllll}
&  &  &  &  & \\
\hline
&  &  &  &  & \\
~~$s$~~&$\gamma_0$~&~$\bar{\gamma}_3^+
$~~&~$\gamma_3^+$~~&$\bar{\gamma}_3^-$~&~~$\gamma_3^-$\\
&  &  &  &  & \\
\hline
&  &  &  &  & \\
2~~~~& 61.1798~~~~& 0.9699~~~~& 2.9033 
~~~~&  1.7599~~~~& 5.2680 \\
2.5    & 61.1878 & 1.5898 & 3.2734 & 2.4612 & 5.0675  \\
2.7349 & 61.1930 & 1.8654 &  3.3020 &  2.7650 & 4.8944 \\ 
3      & 61.1999 & 2.1770 &  3.2783 &  3.1073 & 4.6793  \\
3.5    & 61.2150 & 2.8206 &  3.1856 &  3.8086 & 4.3013 \\
3.5862 & 61.2179 & 2.9445 &  3.1704 &  3.9423 & 4.2448 \\
4      & 61.2325 & 3.6167 &  3.1179 &  4.6608 & 4.0180 \\
&  &  &  &  & \\
\hline
\end{tabular}
\end{center}
\end{table}
The coefficients $\gamma_3^{\pm}$ can be written as a product of the 
universal part $\bar{\gamma}_3^{\pm}$ (table 9) and non-universal factor 
$c_{\nu}^3$ depending on the microscopic parameters of the Hamiltonian 
(the lattice constant $c$, the effective interaction radius $b$ and 
the value $\tilde{\Phi}(0)$ of the interaction potential Fourier 
transform  at $k=0$):
\bee
&&\gamma_3^{\pm}=c_{\nu}^3\bar{\gamma}_3^{\pm}, \nonumber \\
&&\bar{\gamma}_3^{+}=-\gamma^{+}+\bar{f}_{LGR}=-{\gamma}_{+}+\bar{f}_{Ž}
+\bar{f}^{'}, \\
&&\bar{\gamma}_3^{-}=-\gamma^{-}+\bar{\gamma}_{g}+\bar{\gamma}_{\rho}+
\bar{\gamma}_3^{< \sigma >}. \nonumber
\eee

We have for the entropy, internal energy and specific heat of the system
\bee
S &=& 
\left \{
\begin{array}{l}
kN^{'}[s^{(0)}+c_0\tau+u_3^{+}\tau^{1-\alpha}], 
\quad T>T_c, \\
kN^{'}[s^{(0)}-c_0\mid\tau\mid-u_3^{-}\mid\tau\mid^{1-\alpha}], 
\quad T<T_c, 
\end{array}
\right .\nonumber
\eee
\bee
U &=& 
\left \{
\begin{array}{l}
kTN^{'}[\gamma_1+u_1\tau+u_3^{+}\tau^{1-\alpha}], 
\quad T>T_c, \\
kTN^{'}[\gamma_1-u_1\mid\tau\mid-u_3^{-}\mid\tau\mid^{1-\alpha}], 
\quad T<T_c, 
\end{array}
\right . 
\eee
\bee
C &=& 
\left \{
\begin{array}{l}
kN^{'}[c_0+c_3^{+}\tau^{-\alpha}], 
\quad T>T_c, \\
kN^{'}[c_0+c_3^{-}\mid\tau\mid^{-\alpha}], 
\quad T<T_c 
\end{array}
\right .    \nonumber
\eee
with the coefficients given by the relations
\bee
&&s^{(0)}=\gamma_0+\gamma_1, \nonumber \\
&& u_3^{\pm}=c_{\nu}^3\bar{u}_3^{\pm}, \quad \bar{u}_3^{\pm}=
	3\nu\bar{\gamma}_3^{\pm}, \nonumber \\
&&u_1=2\gamma_2+\gamma_1, \\
&&c_3^{\pm}=c_{\nu}^3\bar{c}_3^{\pm},
\quad \bar{c}_3^{\pm}=3\nu(3\nu-1)\bar{\gamma}_3^{\pm}. \nonumber
\eee
The formula for the specific heat can also be rewritten as
\bee
C/kN^{'}=\frac{A^{\pm}}{\alpha}\mid \tau \mid ^{-\alpha}+B^{\pm},\\
A^{\pm}=c_{\nu}^3 \alpha \bar{c}_3^{\pm}, \quad B^{\pm}=c_0. \nonumber
\eee
The plus and minus signs correspond to $T>T_c$ and $T<T_c$, respectively.

The system susceptibility per particle at infinitely small values of the 
external field ${\cal H}$ at $T>T_c ~(\chi=-\frac{1}{N}\frac{\partial ^2 
F_{LGR}}{\partial {\cal H}^2}$, see (2.23)) and $T<T_c$ (see (4.6)) 
is given by
\bee
\chi &=& 
\left \{
\begin{array}{l}
\Gamma^{+}\tau^{-\gamma}\frac{\mu_{B}^2}{\tilde{\Phi}(0)}, 
\quad T>T_c, \\
\Gamma^{-}\mid\tau\mid^{-\gamma}\frac{\mu_{B}^2}{\tilde{\Phi}(0)}, 
\quad T<T_c. 
\end{array}
\right .
\eee
Here (see table 7),
\bee
\Gamma^{+}&=&2c_{\nu}^{-2}\bar{\gamma}_4^{+}, \nonumber \\
\Gamma^{-}&=&c_{\nu}^{-2}\{ 
\frac{10}{3}\frac{\bar{G}^2}{\bar{D}}(-1+\sqrt{1+\frac{3}{2}\frac
{\bar{r}_{\mu_{\tau}+1}\bar{B}\bar{D}}{\bar{G}^2}})(\frac{1}{5}+
\sqrt{1+
\frac{3}{2}\frac{\bar{r}_{\mu_{\tau}+1}\bar{B}\bar{D}}{\bar{G}^2}})- 
\nonumber \\
&-& \bar{r}_{\mu_{\tau}+1}{\bar B}\}^{-1}, \\
\gamma&=&2\nu.\nonumber
\eee

Plots of the temperature dependence of the system free energy $F/N$,  
entropy $S/kN$, specific heat $C/kN$, mean spin moment $<\sigma >$ (4.5), 
susceptibility $\chi$ (5.7) at $s=2, 2.5, 3$ are shown in figures 4-8. 
\begin{figure}
\epsfxsize 125mm
\epsfysize 45mm
\centerline{\epsffile{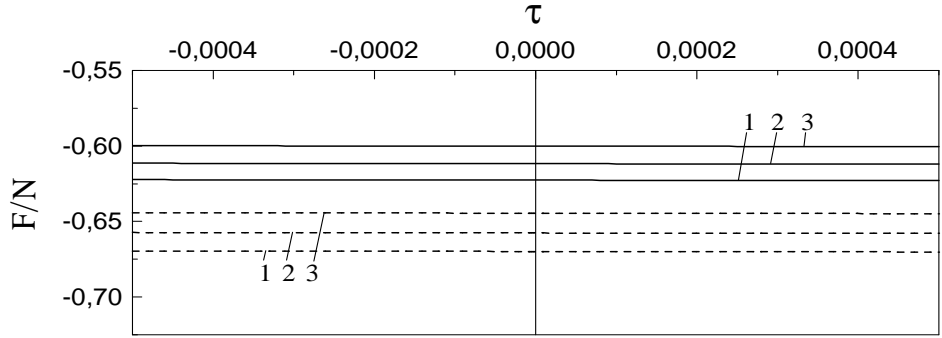}}
\caption {The temperature dependence of the free energy of the system for 
different values of the RG parameter $s$ within the frames of the 
$\rho^6$ model (solid lines). For the comparison we show the 
free energy of the system in the quartic basis distribution 
approximation with allowance for confluent corrections [9] 
(dashed lines). (1) $s$=2, (2) $s$=2.5, (3) $s$=3.}
\end{figure}

\begin{figure}
\epsfxsize 125mm
\epsfysize 60mm
\centerline{\epsffile{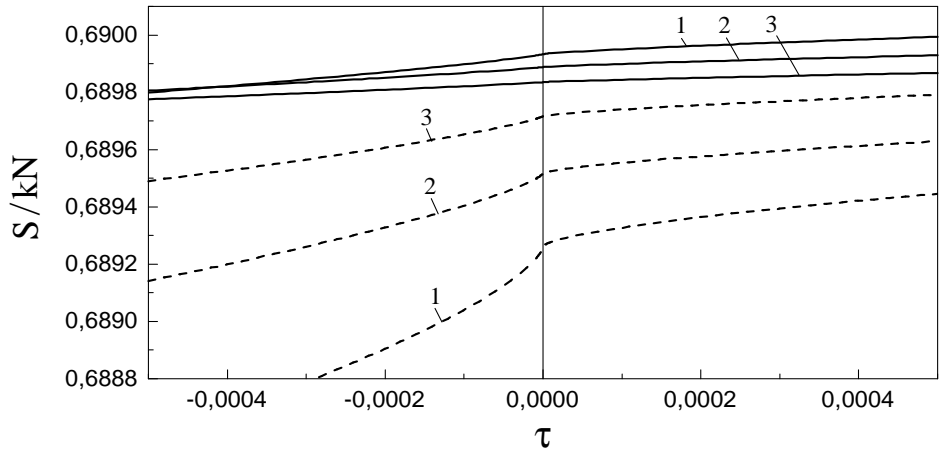}}
\caption {Dependence of the system entropy on $\tau$ (Notations are 
the same as in figure 4).}
\end{figure}

\begin{figure}
\epsfxsize 125mm
\epsfysize 80mm
\centerline{\epsffile{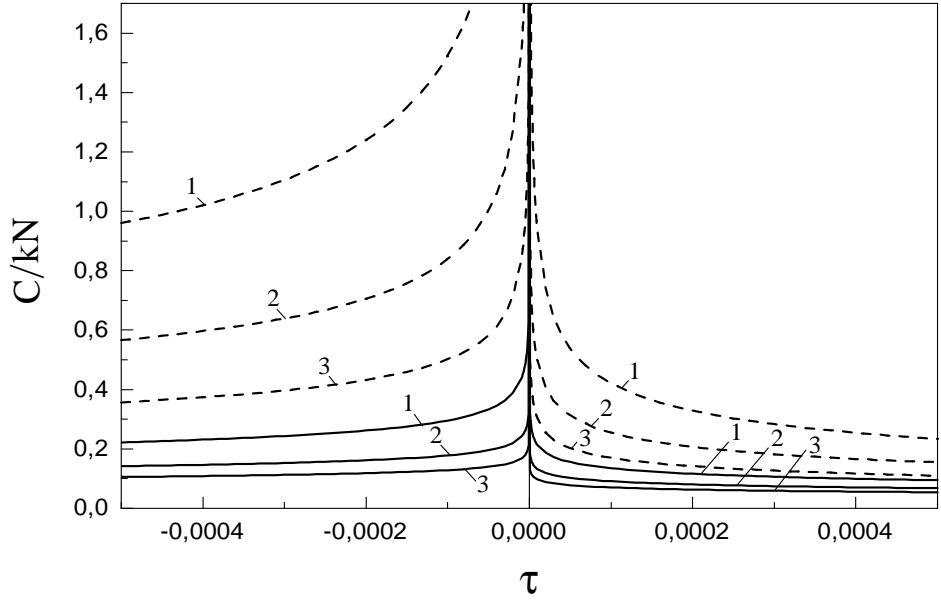}}
\caption {Specific heat of the system (Notations are the same as in figure 
4).}
\end{figure}

\begin{figure}
\epsfxsize 100mm
\epsfysize 60mm
\centerline{\epsffile{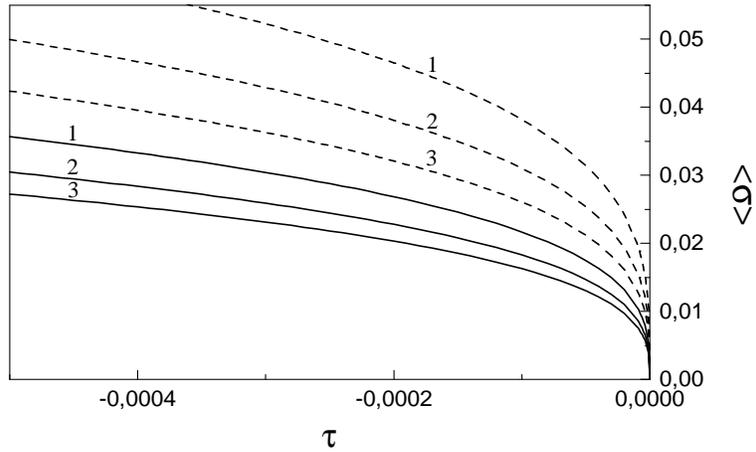}}
\caption {Mean spin moment $<\sigma >$ at $T\leq T_c$ (Notations are the 
same as in figure 4).}
\end{figure}

\begin{figure}
\epsfxsize 125mm
\epsfysize 80mm
\centerline{\epsffile{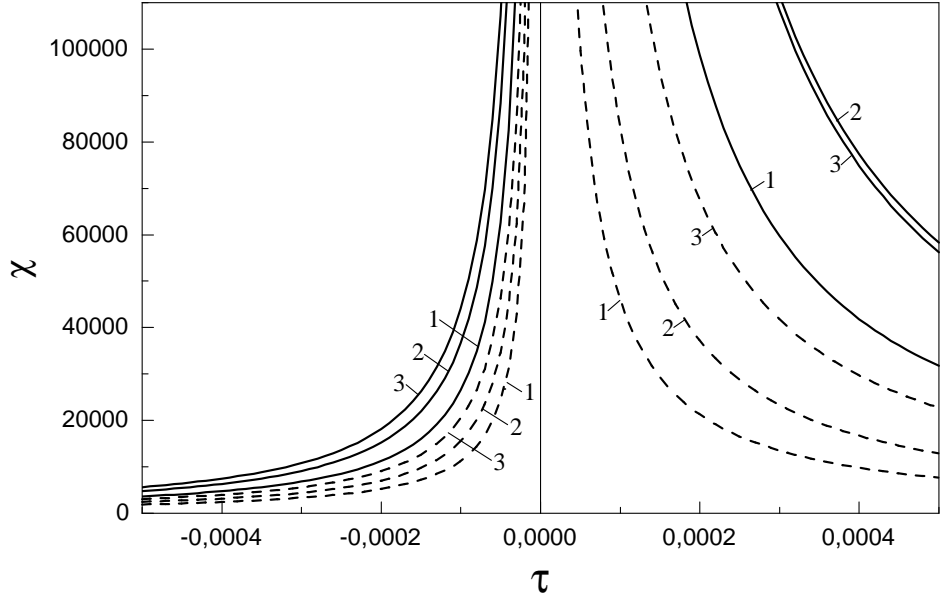}}
\caption {The temperature dependence of the susceptibility $\chi$ 
(Notations are the same as in figure 4).}
\end{figure}
The corresponding plots of the thermodynamic characteristics 
calculated within the $\rho^4$ model, with the confluent corrections [9]
being taken into account, are also given here (dashed curves). 
Comparison of these plots shows that the dependence of the 
thermodynamic functions on the parameter $s$ for the $\rho^6$ model is 
weaker than for the $\rho^4$ model. The dependence of $F/N$ on $s$ for 
these two models is represented in figure 9.
\begin{figure}
\epsfxsize 85mm
\epsfysize 45mm
\centerline{\epsffile{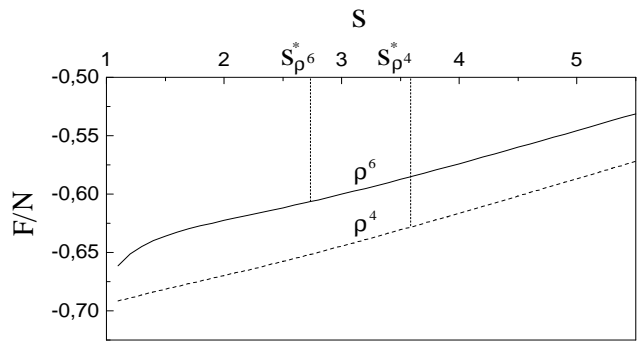}}
\caption {Behaviour of $F/N$ as a function of $s$ for $\rho^4$ and 
$\rho^6$ models.}
\end{figure}

Let us note that the calculations performed are best suited for the 
intermediate values of $s$, close to the quantity $s^{*}$, at which  
$h_n$ turns to zero at the fixed point $(s^{*}$=2.7349 for the $\rho^6$ 
model and $s^{*}$=3.5862 for the $\rho^4$ model). The use of the 
difference form of the RR based on a non-Gaussian measure density 
works especially well for this region of $s$. At small values of 
$s~(s\rightarrow 1)$, some complications arise when the unit element is 
extracted. In this limit, the RR should be represented 
as the perturbation series with respect to the Gaussian distribution 
($h_n$ is large at $s \rightarrow 1$, and expansions in $h_n^{-2}$, 
$\alpha_n$ can be used [19,20]). There also exists an upper limit for $s$. 
At large $s$, one must take into account the correction due to the 
potential averaging [10, 21], which increases with $s$.

The point $s \approx s^{*}$ corresponds to the beginning of the $\nu (s)$
curve stabili\-sation [3]. Table 10 contains the values of the critical 
exponents $\nu, \alpha, \beta, \gamma$, the exponent of the scaling 
correction $\Delta_1=-\ln E_2/\ln E_1$, and the ratios of the critical
amplitudes $A^{+}/A^{-}, \Gamma^{+}/\Gamma^{-}$ and their combinations
$ P=\frac{1}{\alpha}[1-\frac{A^{+}}{A^{-}}]$, $R_c^{+}=A^{+}\Gamma^{+}/
[s_0^3(<\sigma>^{(0)})^2]$ at $s=s^{*}$ for $\rho^4$ and $\rho^6$ models.
These values are in agreement with the data obtained within the field 
theoryapproach (FTA) [22-27] and high-temperature series (HTS) [28-32].
\begin{table}[htb]
\caption{Values of the critical exponents, ratios of the critical 
amplitudes and their combinations for $s=s^{*}$ ($s^{*}$=2.7349 for the 
$\rho^6$ model and $s^{*}$=3.5862 for the $\rho^4$ 
model) obtained by means of the CV method. Data calculated within the field
theory approach (FTA) [22-27] and high-temperature 
series (HTS) [28-32].}
\begin{center}
\begin{tabular}{lllll}
& & & &\\
\hline
\multicolumn{1}{c|}{}&
\multicolumn{2}{|c|}{Model}&
\multicolumn{2}{|c}{Ref. data}\\
\cline{2-5}
\multicolumn{1}{c|}{Quantity}&
\multicolumn{2}{|c|}{~~~~$\rho^4$~~~~~~~~}{~~~~~~$\rho^6$~~~~~~~}&
\multicolumn{2}{|c|}{~~~~~~~~~~~FTA~~~~~~~~~~~}{~~~~~~~~~HTS~~~}\\
\hline
&  &  &  &\\
$\nu$~~~~~ &~~ 0.605~~&~~~~~ 0.637~~~~&~~ 0.630 &0.638 \\
$\alpha$   &~~ 0.185  &~~~~~ 0.088    &~~ 0.110 &0.125  \\
$\beta$    &~~ 0.303  &~~~~~ 0.319    &~~ 0.325 &0.312  \\ 
$\gamma$   &~~ 1.210  &~~~~~ 1.275    &~~ 1.241 &1.250  \\
$\Delta_1$ &~~ 0.463  &~~~~~ 0.525    &~~ 0.498 &0.50 \\
$A^+/A^-$  &~~ 0.435  &~~~~~ 0.675    &~~ 0.54,~~0.48&0.51  \\
$\Gamma^+/\Gamma^-$ &~~ 6.967 &~~~~~ 9.253  &~~~4.77,~~5.12 &5.07\\
$P$        &~~ 3.054  &~~~~~ 3.711    &~~ 3.90, 4.03, 4.2, 4.72 & \\
$R_c^+$    &~~ 0.098  &~~~~~ 0.162    &~~ 0.059, 0.052 &0.059 \\ 
& &  &  & \\
\hline
\end{tabular}
\end{center}
\end{table}

\section*{Conclusions}

\noindent
The method for the calculation of the three-dimensional Ising model 
thermodynamics is developed in the sixfold distribution 
approximation. Both temperature regions above and below the critical value 
of $T_c$ are considered. The main distinguishing feature of the approach is 
the separate allowance for the contributions from the short- and 
long-wave fluctuation phases of the spin moment density to the free energy 
of the system near $T_c$. Within the framework of the $\rho^6$ model, we 
obtained the explicit expressions for the critical amplitudes 
of the thermodynamic functions of the three-dimensional Ising ferromagnet 
and calculated the coefficients of the free energy, the universal 
characteris\-tics (the critical exponents, the ratios of the 
critical amplitudes). Calculation of the free energy, entropy, specific 
heat, mean spin moment, susceptibility was performed for different values 
of $s$. Comparison of the results obtained within $\rho^4$ and $\rho^6$ 
models indicates that the dependence of thermodynamic functions on the 
RG parameter $s$ is weaker for the $\rho^6$ model. 

The values of $s$ close to $s^{*}$ are optimal for the method presented. 
Obtaining analytical expressions for critical amplitudes and 
system thermodynamic characteristics as functions of the Hamiltonian 
microscopic parameters is the advantage of the method developed. The 
leading critical amplitudes for the specific heat and other thermodynamic 
characteristics are represented as a product of the universal 
part, independent of microscopic parameters, and the non-universal factor, 
which depends on these parameters.

\section*{Acknowledgements}

\noindent
This work was supported in part by the Ukrainian State Foundation of 
Fundamental Studies (project No 2.4/173). The authors are pleased to 
express their gratitude to this Foundation.

\end{document}